\documentclass{aa}  

\usepackage{graphicx}
\usepackage{txfonts}
\usepackage{hyperref}
\usepackage{comment}
\hypersetup{colorlinks=true,allcolors=[rgb]{0,0,1}}
\usepackage[dvipsnames]{xcolor}
\usepackage{longtable}
\usepackage{natbib}

\usepackage{ulem}

\newcommand{\HeI}{He\,{\sc i}}
\newcommand{\HeII}{He\,{\sc ii}}

\newcommand{\NII}{N\,{\sc ii}}
\newcommand{\NIII}{N\,{\sc iii}}
\newcommand{\NIV}{N\,{\sc iv}}
\newcommand{\NV}{N\,{\sc v}}

\newcommand{\Teff}{$T_{\rm eff}$}
\newcommand{\logg}{$\log\!g$}
\newcommand{\logQ}{$\log\!Q$}
\newcommand{\He}{$Y_{\rm He}$}
\newcommand{\vsini}{$v \sin\!i$}
\newcommand{\vmac}{$v_{\rm mac}$}
\newcommand{\Nab}{$\epsilon_{\rm N}$}
\newcommand{\kms}{${\rm km}\,{\rm s}^{-1}$}

\begin{document}

   \title{The IACOB project}

\subtitle{XVII. Nitrogen abundances in Galactic O-type stars: further hints for separating binary-interaction products from effectively single stars}
   \author{C. Martínez-Sebastián
          \inst{1,2}
          \and
          G. Holgado
          \inst{1,2}
          \and
          S. Simón-Díaz
          \inst{1,2}
          \and
          F. Martins
          \inst{3}
          \and
          J. Puls
          \inst{4}
          }
   \institute{Instituto de Astrofísica de Canarias, c/Vía Láctea, S/N, E-38205 La Laguna, Tenerife, Spain
   \and
   Departamento de Astrofísica, Universidad de La Laguna, E-38206 La Laguna, Tenerife, Spain
   \and
   LUPM, Université de Montpellier, CNRS, Place Eugène Bataillon, F-34095 Montpellier, France
   \and
   LMU Munich, Universitätssternwarte, Scheinerstrasse 1, 81679 München, Germany
   }
   \date{Received xx,xxxx; accepted xx,xxxx}

  \abstract
  {Growing evidence is revealing the crucial role of binarity in massive star evolution. This additional complexity compounds the uncertainties that still affect single-star evolution models and demands a refinement of the available observational constraints. A first crucial step is to disentangle observed stars that have evolved in isolation from those that have experienced binary interaction.}
  {To investigate the possible evolutionary origins of a sample of 117 Galactic O-type stars with luminosity classes V to III and projected rotational velocities (\vsini) below $\sim$150~\kms.}
  {We mostly focus on surface nitrogen and helium abundances, but also consider other dynamical signatures that may help to distinguish products of binary interaction from effectively single stars. We extend previous quantitative spectroscopic analyses performed within the framework of the IACOB project and obtain N abundance estimates.
  We investigate correlations between these abundances and other stellar parameters, such as \vsini, effective temperature, surface gravity, and He abundance.
  As a reference, we use state-of-the-art predictions from single-star evolution models computed using different physical prescriptions.}
  {We find good agreement between our N abundance estimates and previous determinations based on different analysis methodologies and stellar atmosphere codes. We identify clear differences in the N abundance distributions corresponding to three He abundance regimes, defined as He-low (\He{}\,=\,$N({\rm He})/N({\rm H})$\,$\leq\!0.08$), He-normal ($0.08\!<$\,\He{}\,$\leq\!0.12$), and He-rich (\He{}\,$>\!0.12$). We argue that the abundance estimates for the He-low group, as well as for some additional stars with abnormally low N abundances, are likely spurious determinations. For the He-normal group, the N abundance distribution peaks slightly above the expected birth value and extends up to \Nab{}\,=\,$\log({\rm N/H})+12$\,$\sim$\,8.4~dex. For these stars, we find overall agreement with single-star evolutionary models that include efficient internal mixing and assume moderate-to-low initial rotation ($v_{\rm ini}/v_{\rm crit}\!\lesssim\!0.2$). In contrast, the He-rich group exhibits a bimodal N abundance distribution, with one peak at $\sim$8.1~dex, corresponding to mildly enriched stars, and a second more enriched peak around $\sim$8.5~dex; none of these stars are consistent with predictions from state-of-the-art single-star evolutionary models.}
  {We argue that both N abundance subgroups among He-rich stars are most plausibly explained as binary products. Furthermore, despite N abundance in He-normal stars with LC~IV and V are reproduced by single stellar evolutionary models with efficient mixing models, the same models predict a higher N abundance than observed for stars in this group with LC~III. This indicates that rotational mixing alone is unable to explain the observed distribution of N abundances among stars with normal He abundances.
  A future comprehensive study of surface abundances in O-type supergiants (LC~I and II) and fast rotators, also incorporating abundances of additional elements, is essential to further constrain the evolutionary channels of the most massive stars during the main sequence.}


   \maketitle
%

\section{Introduction}\label{Intro}

\vspace{-5mm}
O-type stars are key to understand the evolution and fate of massive stars. These stellar objects, defined by the presence of \HeI{} and \HeII{} lines in their spectra (\citeauthor{Morgan43} \citeyear{Morgan43} and Sect.~4 in \citeauthor{MaizApellaniz&Negueruela&Caballero26} \citeyear{MaizApellaniz&Negueruela&Caballero26} for a recent review) are located in the upper main-sequence region of the Hertzsprung-Russell diagram, corresponding to initial masses above $\sim\!15~M_{\odot}$ \citep{Humphreys78,Martins+05,Holgado+25}.
As they represent the earliest and longest-lived phase of massive stars, it is essential to understand their evolution during this phase. This information is crucial to the study of a range of astrophysical phenomena in which they play a key role \citep[e.g. core-collapse supernovae, gravitational waves, and galactic and extragalactic evolution][]{Langer12,Mandel&Farmer22,Massey13,Eldridge&Stanway22}.

Nucleosynthesis is the main driver of stellar evolution.
In massive main-sequence stars, the core nuclear processes are thought to be well understood, with hydrogen burning through the CNO cycle as the main energy source. As a result, the He and N abundances increase in the stellar core during this phase, while carbon and --at later stages-- oxygen are depleted. 
However, how these internal abundance changes are reflected at the stellar surface remains an open question even for effectively single stars, as the nature and efficiency of mixing mechanisms are still not fully understood \citep{Martins&Palacios13,Johnston21,Keszthelyi+22}.
In the last decades of the last century, some observational studies began to report surface overabundances of helium and nitrogen in OB stars, challenging the available schema of stellar evolution \citep[e.g.][]{Herrero+92,Gies&Lambert92,Schonberner+88}.

One key process proposed to explain He and N enrichment of the stellar surfaces in O-type stars during the main sequence was rotationally-induced mixing \citep[e.g.,][]{Maeder&Meynet87, Meynet&Maeder00}.
In this framework, stellar rotation can give rise to instabilities that exchange material between the core and the surface, modifying the core size and composition while contaminating the stellar atmosphere with nucleosynthesis material. The adoption of physical prescriptions such as the advecto-diffusive transport \citep[e.g.][]{Maeder&Zahn98}, magnetic dynamo \citep[e.g.][]{Spruit02}, or internal gravity waves \citep[e.g.][]{Talon&Charbonnel03} in the different models, as well as different mixing efficiencies coming from different calibrations leads to different predictions. In this regard, \cite{Martins&Palacios13} presented a systematic comparison of the outcomes from different state-of-the-art evolutionary models, including \cite{Brott+11} and \cite{Ekstrom+12}.

In recent years, several studies have reported a high incidence of multiplicity among Galactic massive stars \citep[e.g.,][]{Chini+12, Sana+12,Sota+14, Barba+17}. Others have claimed a major role of binary interactions in their evolution, revealing a more complex scenario.
In addition to the single star physics, several processes involving the interaction of the various components in a multiple system can take place. For example, tidal forces can regulate the spin rate and affect the rotational mixing \citep{Petrovic+05, deMink+09, deMink+13, Song+13, Sciarini+26}. Moreover, mass transfer can affect the surface abundances of both the donor, which may form a stripped star \citep{Paczynski67, Kippenhahn69, Vanbeveren+98}, and the gainer, which may become enriched in helium and nitrogen \citep{Langer+08, Langer+20, Martinez-Sebastian+25, Jin+26}. 
In fact, \citet{Bolton&Rogers78} first proposed mass transfer as the physical origin of ON stars \citep[O stars with enhanced N absorption lines and deficient carbon and oxygen features;][]{Walborn71,Sota+11}.
Despite subsequent observational studies have found evidence supporting a major role of binarity in this spectral type \citep{Boyajian+05, Li&Howarth20}, their origin remains unclear \citep{Martins+15B}.

In this increasingly complex scenario, observations are essential to constrain stellar evolutionary models and theories. For this reason, the massive-star community has undertaken large spectroscopic surveys covering different regions and metallicities, such as FSMS \citep{Evans+05}, GOSSS \citep{Maiz-Apellaniz+11}, VFTS \citep{Evans+11}, MiMeS \citep{Wade+16}, IACOB \citep{Simon-Diaz+11,Simon-Diaz+20}, OWN \citep{Barba+10,Barba+17}, XShootU \citep{Vink+23}, BLOeM \citep{Shenar+24}, and upcoming projects as WEAVE-SCIP \citep{WEAVE} and 4MIDABLE-LR \citep{4MOST}. Although large samples are required to trace the evolutionary pathways of massive stars, such surveys inevitably comprise a mixture of objects evolved effectively in isolation and others shaped by binary interactions. It is therefore crucial to incorporate additional observational diagnostics that can help distinguish stars evolving effectively as single objects from binary-interaction products; otherwise, the inferred constraints for single stellar evolutionary models may be contaminated by binary products mimicking some of the observed properties in isolated stars.

\cite{Jin+26} have shown that surface abundances --combined with other stellar parameters-- constitute a powerful diagnostic to discriminate between single and binary evolution products. Certain relationships follow strict predictions from nuclear physics that must be satisfied regardless of other stellar processes. An example is the relation between surface enrichment of nitrogen and helium, which \cite{Martinez-Sebastian+25} used to argue that a group of O-type He-rich yet comparatively low N stars are likely binary products.
Moreover, because chemical enrichment depends on the efficiency of internal mixing, studying the surface abundances of stars that evolved in isolation is essential to provide observational constraints for evolutionary models.

The determination of reliable abundances relies on quantitative spectroscopy, in which observed spectra are compared with state-of-the-art atmosphere models. However, different approaches are subject to different limitations depending on the adopted atmospheric code and, more importantly, on the specific comparison techniques employed. To better understand the N abundance in O-type stars, \cite{Rivero-Gonzalez+11,Rivero-Gonzalez+12b,Rivero-Gonzalez+12} developed a detailed N atomic model, implemented it in FASTWIND \citep{Santolaya-Rey+97,Puls+05}, and investigated the formation of N lines in stellar atmospheres. 

Several studies analyzing Galactic O-type stars \citep[e.g.,][]{Martins+15A, Martins+15B, Martins+17, Mahy+22} have used the stellar atmopshere code CMFGEN \citep{Hillier&Miller98}. 
Despite it provides a physically consistent treatment of the stellar atmosphere, the computation of line blanketing by solving the radiative transfer equation of all lines in the co-moving frame makes their models computationally expensive. As a consequence, the sample sizes analyzed in these studies have been effectively limited. Furthermore, the spectral synthesis methodology adopted in these works is sensitive to extrinsic broadening and is also slower than the curve-of-growth method. 

Other studies have used FASTWIND. This code uses an approximation to treat line blanketing that enables a faster model computation, allowing the construction of larger and denser grids.
Based on this code, \cite{Markova+18} used a visual fitting approach to determine abundances in a sample of Galactic O-type stars. This strategy implies some user-subjectivity and is also relatively time-consuming. Therefore, it is only suitable for relatively small samples (30 stars in this study). \cite{Carneiro+19} used a more objective semi-automatic approach, based on the curve-of-growth method --comparing observed and synthetic equivalent widths ($EW$) for multiple lines using an optimized grid of models. It allows for disentangling contributions from lines not explicitly included in the synthetic spectra, making it an ideal technique to use with FASTWIND. Despite this approach enables the analysis of larger samples, it is limited to spectra in which we can measure reliable $EW$s. 


All the aforementioned studies have expanded the available information of N surface abundances in Galactic O-type stars, some of them including also C and O abundances in large samples including all luminosity classes and rotational velocities in O-type stars \citep[e.g.][]{Martins+15A}. Although their different approaches have been shown to be coherent \citep[e.g.][]{Markova+18}, the diversity in methodologies, atmosphere codes, sample selections, and spectra makes it challenging to construct a uniform sample by combining them. Consequently, we still need to increase the number of systematically derived abundances from large, bias-controlled samples. In this context, we aim to contribute to this effort by studying N abundances for a set of 120 Galactic O-type stars observed in the framework of IACOB project (Sect.~\ref{sec_sample}). To achieve this, we apply the curve-of-growth method together with an optimized grid of FASTWIND models (Sect.~\ref{sec_methods}). Our estimated N abundances are consistent with previous studies (Sect.~\ref{sec_results}) and, combined with previously determined spectroscopic parameters and surface He abundances \citep[][]{Holgado+25,Simon-Diaz+26}, allow us to assess the possible evolutionary origins of the sample (Sect.~\ref{sec_discussion}). We found evidence in line with a binary origin of He-rich O-type stars and a bimodal N abundance distribution within these objects. This highlights the interest of future work including fast rotators and supergiants to complement our understanding on massive-star evolution in their early stages (Sect.~\ref{sec_conclusions}). 


\section{Sample}\label{sec_sample}

\begin{figure}[!t]
\centering
\includegraphics[width=.93\hsize,trim={0 15 0 0}]{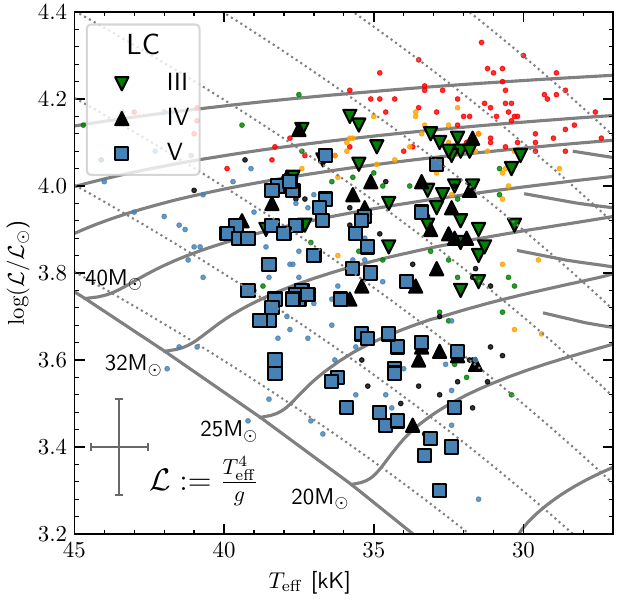}
\vspace*{0mm}
  \caption{Distribution in a spectroscopic HR diagram of the sample of Galactic O-type stars considered in this work (large filled symbols). For reference purposes, the full analyzed IACOB sample of O-type stars included as small size circles; red and yellow symbols for LC~I and II. GENEC evolutionary tracks during the Main Sequence are plotted for $v_{\rm ini}/v_{\rm crit} = 0$ with solid lines. Dotted lines for gravities decreasing from \logg{}\,=\,4.2~dex (left) in steps of $\Delta$\,\logg{}\,=\,0.2~dex.
 \label{fig_sHR}}
\end{figure}

We have selected an optically curated sample in which N abundances can be reliably determined from the total 358 Galactic O-type stars for which the IACOB project has gathered high-quality spectra \citep[last described in][]{Holgado+25}.
Among them, we retained the 225 stars with a very good agreement between the observed spectra and the FASTWIND models for H and He lines. We excluded stars displaying P-Cygni profiles, double-peaked lines, or discrepancies in the fitting between H$\alpha$ and \HeII{}~$\lambda4686$, to avoid possible effects of wind inhomogeneities. As modeling atmospheres with strong winds remains challenging, we restricted our sample to 210 stars with \logQ\,$\leq\!-12.7$ \citep[wind-strength parameter, $Q=\dot{M}/({\rm M}_{\odot}\;{\rm yr^{-1}})/(v_{\infty}/({\rm km\;s^{-1}})R/({\rm R}_{\odot}))^{3/2}$; first described in][]{Puls+96}. Given the extremely low statistics resulting from the application of this criteria for stars with luminosity classes (LC) I and II (i.e. supergiants and bright giants), we decided to limit ourselves to LC~III, IV, and V.
We imposed two additional criteria on the signal-to-noise ratio ($S/N > 50$) and the apparent visual magnitude of the stars ($m_{\rm V} \leq 10$) to ensure the quality of the spectra. This results in a sample of 145 stars.
Our sample contains SB1 systems while is curated to exclude detected SB2 binaries. Nonetheless, some stars may still harbor undetected companions\footnote{We note that our binarity classification differs from that adopted in other studies \citep[e.g.][]{Mahy+22}. We refer the reader to Appendix~\ref{app_dilution} for further details.} \citep[][]{Simon-Diaz+26}.

After this initial selection, we conducted a visual pre-analysis of the resulting subset to ensure the availability of diagnostic lines for the determination of N abundances. We inspected two strong lines per ion (\NII{} 3994.99 and 6610.56~$\AA$, \NIII{} 4379.11 and 4514.86~$\AA$, \NIV{} 4057.76 and 6380.77~$\AA$\footnote{We note the presence of a strong, overlapping diffuse interstellar band.}, and \NV{} 4603.73 and 4619.98~$\AA$) to assess whether they are reliably detected in the available spectra (further details in Appendix~\ref{app_selection}). Because \NIII{} lines are the most prevalent across the sample and are modeled with high reliability, we focused our abundance analysis on stars showing strong \NIII{} features (effective temperature \Teff{}\,$\leq 40$~kK). Finally, to ensure an accurate $EW$ measurement, we limited our sample to stars with low projected rotational velocities (\vsini{}\,$\leq\!150$~\kms{}).

Figure~\ref{fig_sHR} depicts the distribution of the final sample of 117 stars in the spectroscopic HR diagram \citep[sHRD,][]{Langer&Kudritzki14} with large filled symbols. The shape and color of each marker indicates the luminosity class of each star. For reference, we include the full set of 358 O-type stars analyzed in \cite{Holgado+25} and the non-rotating evolutionary tracks computed by \citet{Ekstrom+12} using the code GENEC\footnote{https://www.unige.ch/sciences/astro/evolution/en/database}.

The distribution of the analyzed stars is approximately uniform for $\log(\mathcal{L}/\mathcal{L_{\odot}}) < 4.2$ and $30 \leq T_{\rm eff} \leq 40$~kK, roughly corresponding to luminosity classes III, IV, and V. Any discussion beyond this regime remains speculative and would require additional work.

\section{Data and methods}\label{sec_methods}

The analyzed spectra have been gathered in the framework of the IACOB project, with resolving power in the range between 25\,000 and 85\,000 \citep{Simon-Diaz+11,Simon-Diaz+20}. In each case, we use the spectra from the IACOB database with the best signal-to-noise ratio, with $\sim$80\% of them having a $S/N$\,$\gtrsim$\,100.

For each star, we benefited from information about its effective temperature (\Teff), surface gravity (\logg{}), line-broadening parameters (\vsini\ and macroturbulent velocity, \vmac), and wind-strength parameter (\logQ), and surface He abundance (\He{}\,=\,$N({\rm He})/N({\rm H})$) as provided in \cite{Holgado+25} and \cite{Simon-Diaz+26}\footnote{While both works rely on a similar quantitative spectroscopic analysis strategy, for common stars we prefer the most updated values of \cite{Simon-Diaz+26}.}. We consider these parameters well determined and keep them fixed during the N abundance analysis. Nonetheless, we finetune some values when necessary (e.g. to correct inconsistencies in the nitrogen ionization balance).

For the estimation of the N abundance (\Nab\,=\,log(N/H)\,+\,12), we follow an analysis strategy similar to \citet{Carneiro+19}, based on the curve-of-growth method. We use the $EW$s of a suitable set of diagnostic lines (Table~\ref{tab_lines}) to determine N abundances by comparing them with those predicted by the closest reference model from our FASTWIND grid (Sect.~\ref{grid}). This approach is largely insensitive to extrinsic broadening mechanisms (i.e. \vsini{} and \vmac{}) and to radial-velocity corrections. {\iffalse, since these effects preserve the $EW$ of the intrinsic spectrum.\fi}Consequently, it provides a more reliable determination of the microturbulent velocity ($\xi$). In addition, this method allows for a faster reanalysis once the $EW$s are measured.

\subsection{Mesurement of equivalent widths}

We measure the $EW$s by fitting the observed lines with a modeled profile and integrating the resulting fits. We prefer this approach over direct integration because it allows us to treat blended lines separately through multiple-line fitting. Each profile is based on a Gaussian with an $EW$ corresponding to the atomic profile and a full-width-half-maximum corresponding to the instrumental profile and a free component accounting for the intrinsic broadening (i.e. microturbulence). This is convolved with two broadening functions representing rotation and macroturbulence effects. To achieve the best possible fit, we let \vsini{}, \vmac{}, and the line center $\lambda_0$ vary as semi-free parameters. We treat the line depth and intrinsic broadening as free parameters, and fix the continuum level at unity.

For each star, we try to measure a total of 24 \NII{}, 12 \NIII{}, and 3 \NIV{} lines\footnote{Some lines belong to the same multiplet (e.g., 4041.31 and 4043.53 in \NII{}, or 4510.88, 4514.85, 4518.18, 4523.56, and 4534.58 in \NIII{}) and are used to constrain $\xi$, as discussed later.} (Table~\ref{tab_lines}).
Noteworthy, we never use all the listed lines, as their detection depends of the specific parameters of the stars, but we use as many as possible in each case.
To account for the main source of uncertainty in the observed $EW$ (i.e. the continuum normalization), we decided to fit each line using both the original and a locally renormalized continuum. This allows us to estimate the $EW$ uncertainty ($\Delta EW$) and to select the optimal normalization as the one that minimizes the mean residual between the observed and fitted profiles.

\begin{table}
    \centering
    \begin{tabular}{lllll}
        \hline\hline
        Ion & \multicolumn{4}{c}{Wavelength ($\AA$)}  \\
        \hline
        \NII{} & 3995.85 & 4035.08 & 4041.31 & 4043.53 \\
        & 4447.03 & 4601.47 & 4607.16 & 4613.87\\
        & 4621.39 & 4803.29 & 5001.47 & 5005.15\\
        & 5007.33 & 5025.66 & 5045.10 & 5666.63\\
        & 5676.01 & 5679.55 & 5710.77 & \\
        \hline
        \NIII{} & 3998.65 & 4003.65 & 4097.35* & 4195.76*\\
        & 4379.11 & 4510.88 & 4514.85 & 4518.18\\
        & 4523.56 & 4534.58 & 4634.12* & 4640.64*\\
        & 4641.85* & & &\\
        \hline
        \NIV{} & 4057.76 & 5205.15 & 6380.77 & \\
        \hline
    \end{tabular}
    \caption{Diagnostic lines used in this work to derive N abundances. Lines marked with an asterisk are only used for qualitative inspection of the final fit (hence not included in the abundance determination).} 
    \label{tab_lines}
\end{table}

\subsection{Grid of FASTWIND models}\label{grid}

We constructed a grid of synthetic spectra with FASTWIND v10.6.5 \citep{Santolaya-Rey+97, Puls+05, Rivero-Gonzalez+12b}, optimized to analyze N abundances in O-type stars \cite[previously used in][]{Martinez-Sebastian+25}. We adopt parameter ranges similar to those of the grid described in \cite{Simon-Diaz+26}, but including C, N, and O as explicit elements and only three values of \He{}\footnote{In this work we use the He abundance computed in \cite{Simon-Diaz+26}.}.
We fix the exponent of the wind velocity law to $\beta = 1$ and neglect clumping --consistent to first order with our sample selection criteria. We run the models assuming solar metallicity \citep[][]{Asplund+09} and adopt the same atomic models as \citet{Carneiro+19}, including the nitrogen model atom developed in \citet{Rivero-Gonzalez+11, Rivero-Gonzalez+12b, Rivero-Gonzalez+12}. Table~\ref{tab_grid_param} summarizes the parameter ranges. 
As a sanity check, we confirm that all \ion{H}{i}, \ion{He}{i} and \ion{He}{ii} line profiles are consistent between our grid and that of \cite{Simon-Diaz+26}.


\begin{table}
    \centering
    \vspace{-0mm}
    \begin{tabular}{ll}
        \hline\hline
    Parameter & Ranges or specific values\\
    \hline
    \Teff{}~[kK] & [25\,--\,60] (Step: 1) \\
    \logg{}~[dex] & [2.7\,--\,4.5] (Step: 0.1) \\
    \He{} & 0.08, 0.1, 0.15 \\
    $\xi$~[\kms{}] & 1, 3, 5, 7, 9, 11, 15, 20, 25, 30\\
    \logQ{} & -12.1, -12.5, -12.7, -13, -13.5, -14 \\
    $\epsilon_{\rm C}$; $\epsilon_{\rm N}$; $\epsilon_{\rm O}$ & 7.80; 8.35; 8.75 \\ ~[dex] & ($\pm0.05$; $\pm0.10$; $\pm0.15$; $+0.20$; $\pm0.30$;\\ 
    & +0.40; +0.50; $\pm0.60$; +0.90; +1.20)\\
    \hline
    \end{tabular}
    \caption{Parameter coverage of the FASTWIND grid used in this work. \label{tab_grid_param}}
    \vspace{-5mm}
\end{table}

\subsection{Abundance and microturbulence determination}\label{ab_determination}

We determine  consistently and simultaneously $\xi$ and \Nab{}.
We use the measured $EW_{\rm l,obs}$ to map the parameter space of \Nab{} and $\xi$ for each line using a $\chi^2$:
$$\chi^2_{\rm l}=\left(  \frac{EW_{\rm l,obs}-EW_{{\rm l,mod}}}{\Delta EW_{\rm l,obs}}\right)^2.$$

To refine our mapping, we interpolate the model equivalent widths ($EW_{\rm mod}$) in steps of $\Delta\xi\!=\!1$~\kms{} and $\Delta\epsilon_{\rm N}\!=\!0.01$~dex.

\begin{figure}[!t]
\includegraphics[width=0.95\hsize,trim={0 1 0 30, clip}]{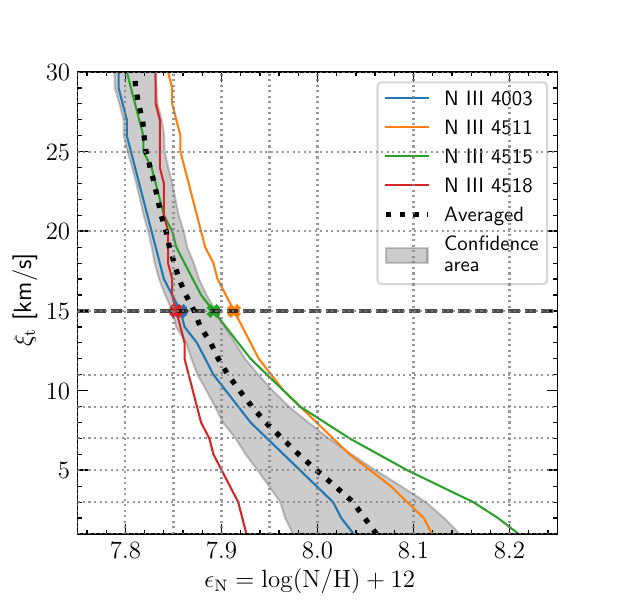}
  \caption{Example of the mapping of N abundance as a function of microturbulence. Colored lines show the N abundance that minimizes $M_{\rm l}$ as a function of microturbulence for each of the analyzed lines in HD~206~183. 
  The horizontal dashed line marks the $\xi$ value determined with the \NIII{} multiplet. The colored X marker corresponds to the abundance of each line at this microturbulence.
  The dotted line indicates the average abundance, with the gray area representing the 1-$\sigma$ dispersion around this value. The dotted gray lines in the background denote the grid steps.}
 \label{fig_maplines}
\end{figure}

For each line, we select the abundance that minimizes $\chi^2_{\rm l}$ for a given $\xi$. In Fig.~\ref{fig_maplines} we show an example of the diagnostic diagram we use with all available \NIII{} lines for the O9.5\,IV-V star HD~206~183 (the same star is used in subsequent figures within this section). 
For each ion, we then compute a weighted average minimum from the different lines with a factor $w_{\rm l}$ that average the different lines from the same multiplet to avoid giving extra weight to transitions between the same levels ($\sum\limits_{\text{l in mult}}\!w_{\rm l}\!=\!1$) (dotted line in Fig.~\ref{fig_maplines}).

We determine $\xi$ using lines from the same multiplet whenever possible. If so, we identify the $\xi$ value that minimizes the abundance scatter among the lines of the same multiplet. We prefer this approach over the use of all the lines to minimize spurious results due to the scatter of different lines caused by the normalization of different regions. With this approach, we can ensure a consistent normalization over the used lines.
We assume a single $\xi$ for all nitrogen ions, so this determination is valid for the element as a whole. In the example of Fig.~\ref{fig_maplines}, the resulting value is shown with the horizontal dashed line at $\xi = 7$~\kms{}.
When this strategy is unfeasible, we estimate $\xi$ by comparison with stars of similar LC and \logg{} for which the microturbulence is well constrained.

\begin{figure}[!t]
\includegraphics[width=0.9\hsize,trim={0 1 0 0}]{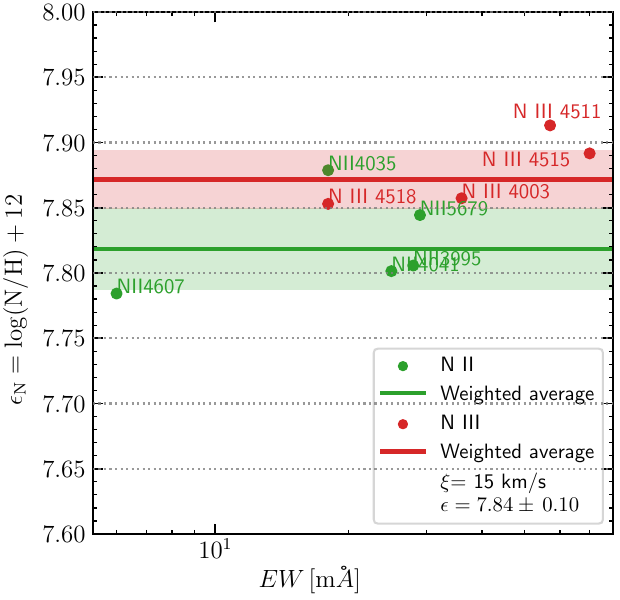}
\vspace*{-3mm}
  \caption{Example of the diagnostic plot used to correct possible systematic effects in the abundance determination. It shows abundances of the different analyzed lines in HD~206~183 for a fixed microturbulence velocity. Green and red symbols correspond to \NII{} and \NIII{} lines, respectively. Solid lines in the same colors indicate the average abundance for each ion, with the shaded areas being their 1-$\sigma$ confidence intervals.\label{fig_ab_EW}}
\end{figure}

\begin{figure*}[!t]
\begin{center}
\includegraphics[width=.93\hsize,trim={0 1 0 0}]{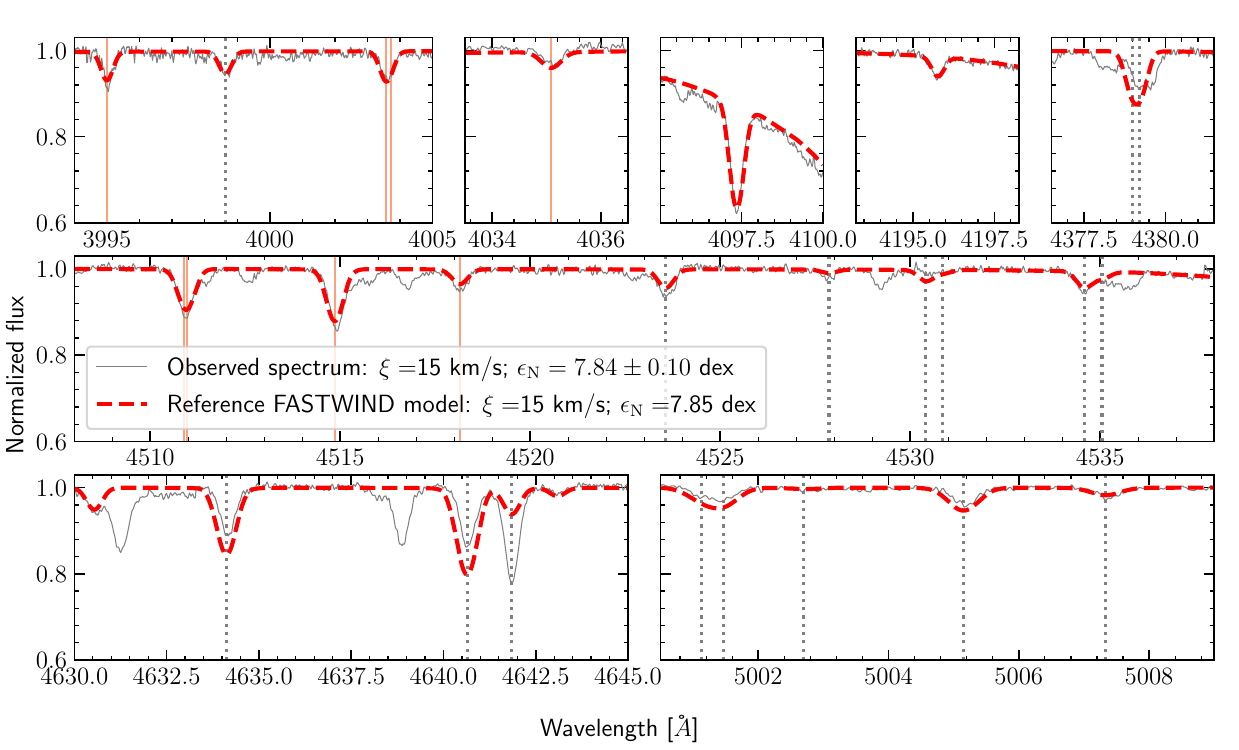}
\vspace*{0mm}
  \caption{Example of the evaluation of the N abundance fit in selected spectral regions. Orange vertical lines denote the nitrogen transitions employed in the abundance determination; dotted gray vertical lines indicate other N lines present in the model but excluded from the analysis. Note that at the temperature of this object, \NIII{}~4642 is dominated by an overlapping O\,{\sc ii} line \citep{Rivero-Gonzalez+11}.\label{fig_quality}}
\end{center}
\end{figure*}

In a subsequent step, we select the abundance for each line corresponding to the adopted $\xi$ (marked with X in Fig.~\ref{fig_maplines}), and determine the surface N abundance as the total average of all lines. We estimate the uncertainty as the standard deviation of all studied lines, adopting a minimum value of $\Delta$\Nab{}\,$=\!0.15$~dex (justified in Appendix~\ref{app_errors}). We then plot the abundance of each line as a function of $EW_{\rm obs}$ to visually identify possible inconsistencies (Fig.~\ref{fig_ab_EW}).
If we detect systematic discrepancies in resulting line-by-line abundances that can be linked to a fundamental parameter (e.g., an incorrect ionization balance), we repeat the analysis after adjusting the reference model.
In a final step, we assess the quality of the analysis in a visual review by overplotting the closest synthetic spectra from the grid onto the observations for key diagnostic lines (Fig.~\ref{fig_quality}). 

We note that the parallel determination of microturbulence and N abundance in this work may lead to $\xi$ values that differ from those derived from He in \cite{Simon-Diaz+26}. To our knowledge, there is no evidence that the microturbulence associated with nitrogen and helium in O-type stars must be the same. Nevertheless, we consider the impact that $\xi$ may have on the He abundance determination \citep[e.g.][]{McErlean98,Villamariz&Herrero00,Howarth&Smith01,Massey+13,Markova+20}. Given the relevance of this effect for the following discussion, we repeated the He abundance analysis using the microturbulence derived in this work. Comparing both determinations, the results remain consistent within the uncertainties and do not affect the main conclusions of this study (Appendix~\ref{app_He}). Without any further clear physical argument to favor one set of microturbulence values over the other, we adopt our determinations, as they ensure internal consistency with the N abundance analysis.

\section{Results}\label{Results}\label{sec_results}

Table~\ref{tab_results} presents the N abundance estimates for the 117 analyzed stars, together with additional parameters relevant to our study. We also include the most up-to-date spectral classifications from GOSSS \citep{Sota+11, Sota+14, Maiz-Apellaniz+16}, as well as the runaway status from \citet{Carretero-Castrillo+23,Carretero-Castrillo+25} and private communication. We organize the table into two groups according to luminosity class (IV+V and III), each sorted by spectral type. 
With respect to the quality of the results, we have 85 stars with a very good fitting of the diagnostic lines and 34 with a compromise solution, in which some lines would require different abundances. Nonetheless, all of them have good fittings considering uncertainties.

\subsection{Comparison with literature} \label{sec_bib_comparison}

This study includes 28 stars previously analyzed in the literature using different stellar atmosphere codes and methodologies. \citet{Martins+15A, Martins+15B, Martins+17} and \citet{Mahy+22} used CMFGEN models and a spectral synthesis methodology. They assumed a microturbulence varying from 10~\kms{} at the photosphere to 10\% of the terminal velocity at the outer boundary of the atmosphere. \citet{Aschenbrenner+23} applied a hybrid non-LTE method, performing non-LTE spectral synthesis with DETAIL and SURFACE on LTE model atmospheres computed with ATLAS12, and derived abundances by fitting individual element lines. \cite{Markova+18} and \cite{Carneiro+19} used FASTWIND models but adopted different analysis strategies. Namely, \citet{Markova+18} derived N abundances by fitting a set of nine lines with fixed microturbulent velocities of $\xi=10$~\kms{} for cooler and $\xi=15$~\kms{} for hotter stars, whereas \citet{Carneiro+19} used the curve-of-growth.
We use the results of N surface abundances of these studies to assess the consistency of abundance determinations obtained by different authors, with  different model atmosphere codes, and analysis techniques.
As a sanity check, we compared the observed spectra with the closest models from our grid using both our results and those from previous publications for all common stars. In general, for our grid, our results provide a better match to the diagnostic lines analyzed in this work. This confirms that our tools and methodology introduce no significant systematics.

\begin{figure*}[!t]
\centering
\includegraphics[width=.93\hsize,trim={0 1 0 0}]{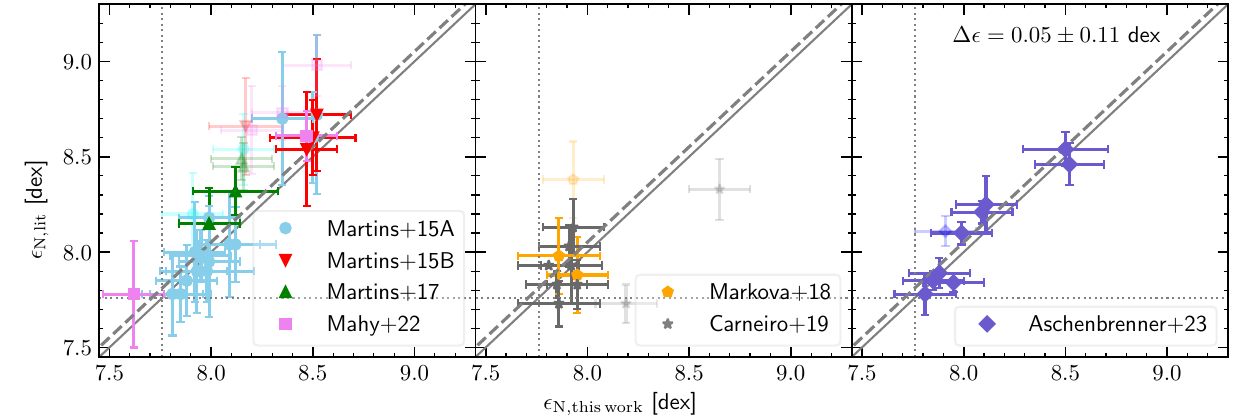}
  \caption{Comparison of N abundances derived in this work with values from previous studies. Different panels for different abundance analysis methods and atmosphere codes (see text).
  Transparent symbols correspond to stars with discrepant comparisons (beyond uncertainties). Vertical and horizontal dotted lines indicate the CAS values at the solar position \citep{Nieva&Przybilla12}. The solid and dashed diagonal lines represent, respectively, the one-to-one relation and the mean relation respect to literature (obtained by fitting the consistent stars considering a fixed slope of unity), with the offset between both relations indicated at the top of the figure.
 \label{Fig_ab_comp}}
\end{figure*}

In Fig.~\ref{Fig_ab_comp}, we compare our abundances with previous determinations.
Different colors and symbols identify the reference studies. We group the various atmosphere codes and abundance-determination approaches --summarized in the previous paragraph-- into separate panels.
We use solid symbols for stars whose \Nab{} values agree within the uncertainties and transparent symbols for inconsistent determinations. For reference, we include the cosmic abundance standard (CAS) for nitrogen from \citet{Nieva&Przybilla12}, hereof considered the reference baseline abundance for our sample. 

The comparison sample spans in \Nab{} values between 7.6 and 8.5~dex, which broadly overlaps with our covered range. The distribution, however, is not uniform: 21 stars cluster between 7.7\,$<$\,\Nab{}\,$<$\,8.2~dex, while only seven\footnote{We warn the reader that some stars appear more than once in Fig.~\ref{Fig_ab_comp} because they were analyzed in multiple studies.} have \Nab{}\,$\geq$\,8.2~dex.
Despite this uneven sampling, the number of comparable stars is sufficient across the full range to draw meaningful conclusions.
We find overall good agreement with previous studies, with a small systematic shift toward lower abundances in our results. The offset amounts to 0.06~dex, below the typical abundance uncertainty (0.15~dex) and the intrinsic scatter in the comparison (0.11~dex).

In the first panel (literature abundances derived with spectral synthesis using CMFGEN), we find good agreement for stars with \Nab{}\,$\lesssim\!8.1$~dex. At higher abundances (\Nab{}\,$\gtrsim\!8.1$~dex), our results show a slight systematic offset toward lower values ($\Delta$\Nab{}\,$\sim\!0.1$~dex). However, the relatively large uncertainties in this regime make most of the common stars consistent within errors across the different studies.

In the second panel, we include the two studies that use the same atmospheric code as in this work. In the case of \citet{Carneiro+19} --whose methodology most closely resembles ours-- we find very good agreement for stars with \Nab{}\,$<\!8.15$~dex. However, two objects (HD~12~993 and HD~303~311, shown as semi-transparent gray star markers in the middle-right and lower-left regions of Fig.~\ref{Fig_ab_comp}, respectively) have significantly higher abundances in our analysis. For the O6.5V((f)) Nstr star HD~12~993, our fit reproduces the observed spectrum more accurately, while for the O6V((f))z star HD~303~311 we obtain a compromise solution, with some lines fitted better and others worse. This highlights that the line selection for abundance determinations directly affects the results.
We share only three stars in common with \citet{Markova+18}. Two display good agreement between both studies, while for the third one they obtain a higher abundance. Given the limited statistics and the different methodologies -- ours being more systematic -- we do not consider this significant.

Finally, in the third panel, we find very good agreement with \citet{Aschenbrenner+23}, with no significant systematic differences across the full range of abundances covered.

\subsection{Overall distribution}\label{sec_GeneralResults}

\begin{figure}[!t]
\centering
\includegraphics[width=.93\hsize,trim={0 1 0 0}]{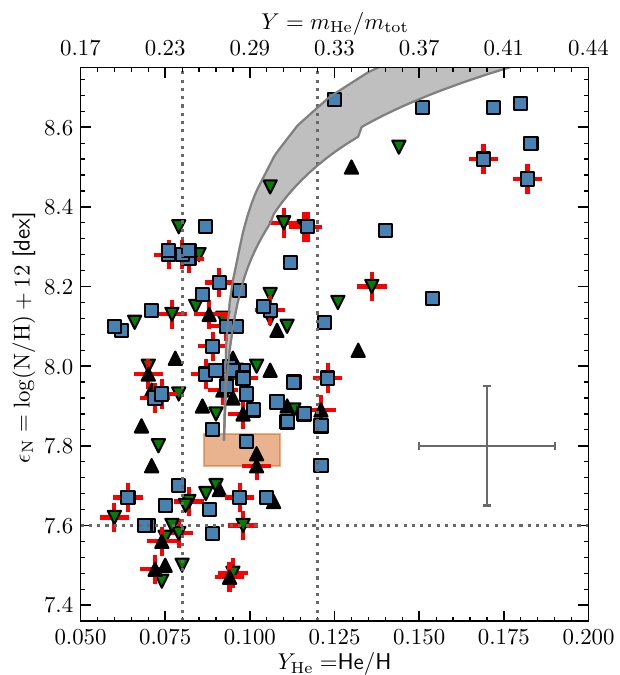}
  \caption{Distribution of the sample in the N against He abundance plane (cf. Fig.~1 in \citealt{Martinez-Sebastian+25}). The gray area marks the region covered by \citet{Ekstrom+12} models for 20\,--\,60~$M_{\odot}$ and $v_{\rm ini}/v_{\rm crit}=0.2$\,--\,0.4 during the MS; the light brown area indicates CAS values from \citet{Nieva&Przybilla12}. Symbols as in Fig.~\ref{fig_sHR}; red cross on the back to mark SB1 systems. Vertical dotted lines mark the He low and rich limits (\He{}$\leq0.08$ and \He{}$>0.12$); the horizontal dotted line separates N-low stars. In upper X-axis, He by mass; in lower X-axis, He by number
 \label{fig_letter}}
\end{figure}

In Fig.~\ref{fig_letter}, we present the relation between N and He abundances \cite[the latter coming from][]{Simon-Diaz+26} for the analyzed stars, using the same color and symbol coding as in Fig.~\ref{fig_sHR}. As a reference, we include predictions from single-star evolutionary models computed with GENEC for initial masses between 20 and 60~$M_{\odot}$ and rotational velocities $v_{\rm ini}/v_{\rm crit}$\,=\,0.2\,--\,0.4 \citep[gray shaded region,][and private communication]{Ekstrom+12}. We also mark the CAS values for the solar neighborhood from \citet{Nieva&Przybilla12} in light brown. The cross in the bottom right corner indicates the standard uncertainties for both quantities. Following \cite{Simon-Diaz+26}, we divide the sample into three groups based on their He abundances divided by the vertical dotted lines.
These groups --defined as He-low, He-normal, and He-rich by \cite{Simon-Diaz+26}-- correspond to \He{}\,$\leq\!0.08$, $0.08\!<$\,\He{}\,$\leq\!0.12$,and \He{}\,$>\!0.12$ by number (i.e. $Y\!\leq\!0.24$, $0.24\!<\!Y\!\leq\!0.32$, and $Y\!>\!0.32$ by mass), respectively. 
The horizontal dotted line identifies the threshold in N abundance below which the estimated abundance is clearly incompatible with the CAS baseline, even when we take into account the typical uncertainties of the determinations ($\Delta$\Nab{}$\sim$0.15~dex; Sect.~\ref{sec_lowN}).

From a theoretical standpoint, we expect a strong correlation between \He{} and \Nab{} for single stars; significant surface helium enrichment (i.e. \He{}$\geq0.12$) should occur only after it has been clearly enriched in nitrogen (i.e. \Nab\,$\gtrsim\!8.5$~dex for CAS initial abundances). This strong correlation was observed in LMC stars by \cite{Rivero-Gonzalez+12} and \cite{Grin+17}, and also including the SMC in \cite{Martins+24}. In our sample, most stars ($\sim$65\%) agree with the predictions of single stellar evolution models within their uncertainties in this projection. However, several outliers remain. Namely, stars with (i) unexpectedly low He abundances \citep[left region; discussed in][and later in this work]{Simon-Diaz+26}; (ii) N abundances below the CAS reference (under the horizontal dotted line); and (iii) relatively low N enrichment (\Nab\,$\lesssim$\,8.4~dex) compared to their clear He enhancement (\He{}\,$\gtrsim$\,0.12).

Thanks to the larger size of the analyzed sample compared to previous works, \cite{Martinez-Sebastian+25} were able to identify, for the first time, a non-negligible group of stars with high He but mild N enrichment (corresponding to the third group of outliers previously indicated). Notably, the sample considered in the aforementioned work differed from the one in this study in two key aspects. 
First, the selection criteria for targets to be considered in our previous study were less restrictive. In particular, we included stars classified as LC I and II, and some other cases which has been discarded for the present work (Sect.~\ref{sec_sample}).
Second, the group identified as He-low --which comprises $\sim$28\% of the current sample--  was excluded in our previous work, since their low He abundances lacked a clear physical explanation and their omission did not impact the main conclusions of that study.
In the present work, we identify 11 out of the 19 stars with \He{}\,$\geq\!0.12$ with N abundances under the expected values according to \cite{Ekstrom+12} single stellar evolutionary models even considering their uncertainties (i.e. about 60\% of the He-rich subsample). Notably, this fraction is somewhat lower than the $\sim$80\% reported in \citet{Martinez-Sebastian+25}. The difference arises primarily from the reanalysis we have performed, which incorporates additional diagnostic lines and yields slightly revised abundances. While these variations remain within the expected uncertainty range, they reconcile part of the He-rich population with predictions from the considered models in this diagram. Nonetheless, the presence and frequency of outliers among the He-rich stars remain statistically significant and in line with the interpretation proposed in \cite{Martinez-Sebastian+25} that binary interaction is a likely origin for these objects.

\begin{figure}[!t]
\centering
\includegraphics[width=.93\hsize,trim={0 1 0 0}]{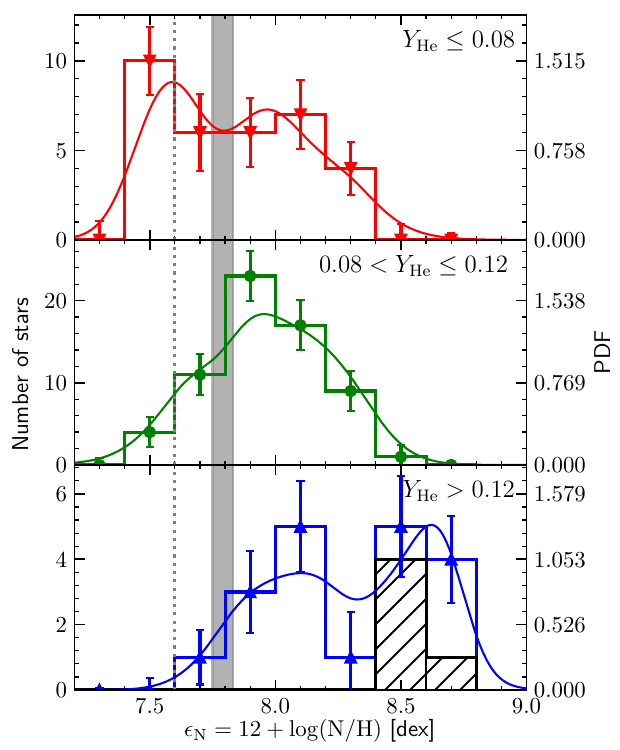}
  \caption{Observational histograms (steps) and probability density estimates (solid lines) of \Nab{} for the sample, separated --from top to bottom-- into He-low, He-normal, and He-rich (red, green, and blue, respectively). The CAS \Nab{} value from \cite{Nieva&Przybilla12} is indicated by the shaded vertical band. Vertical dotted lines mark the threshold separating N-low stars. Black dashed bins highlight ON stars.
 \label{fig_histNitrogen_byHe}}
\end{figure}

Complementary to Fig.~\ref{fig_letter}, Fig.~\ref{fig_histNitrogen_byHe} presents the observed N-abundance distributions for stars in the three regimes of He abundances (Sect.~\ref{sec_discussion}). From top to bottom, the panels correspond to the He-low, He-normal, and He-rich subsamples. Each panel shows the histogram of the sample, including the uncertainty of each bin, with the number of sources indicated on the left y-axis. The empirical probability density function -- computed by summing the individual density functions of all measurements -- is overplotted, with its scale shown on the right y-axis. The CAS value is indicated by the gray vertical band in all panels. Vertical dotted lines separate the N-low regime. Bins corresponding to ON stars are highlighted in black. 

The three distributions differ. An Anderson–Darling test \citep{AndersonDarling,Scholz+87} between the He-low and He-normal groups, and between the He-normal and He-rich groups, yields $p$-values of $5\cdot10^{-5}$ and $3\cdot10^{-4}$, respectively. Therefore, we can reject the null hypothesis that the samples are drawn from the same parent distribution. Consequently, we discuss each subsample separately.

Roughly $\sim$27\% of the analyzed sample (32 stars) show \He{}\,$\leq\!0.08$, lower than the expected birth abundance even within uncertainties (left region in Fig.~\ref{fig_letter}). \cite{Simon-Diaz+26} propose that the estimated abundances for this group of stars is likely a consequence of the impact of additional light from a nearby source contaminating the analyzed spectra (also Sect.~\ref{sec_lowN} and Appendix~\ref{app_dilution}).
This group has an SB1 incidence of $\sim$48\% and spans \Nab{} values from $\sim$7.4 to $\sim$8.4~dex. Its N abundance distribution is asymmetric, with a global maximum at \Nab{}$\,\sim\!7.6$~dex  --significantly below the CAS value for nitrogen--  and a secondary peak around \Nab{}$\,\sim\!7.95$~dex. A total of 15 stars have \Nab{} below CAS, 11 of which remain inconsistent even within uncertainties ($\sim$34\% of the subset against $\sim$6\% in the rest of the sample).

The majority of the analyzed sample ($\sim$56\%; 65 stars) show normal He abundances. This group has an SB1 incidence of $\sim$42\% and spans \Nab{} values from $\sim$7.4~dex to $\sim$8.8~dex. The N abundance distribution is more symmetric than that of the He-low sample, with a maximum around \Nab{}\,$\sim\!7.9$~dex, coinciding with the secondary peak of the He-low group. The decline at \Nab{} values below CAS is steeper than at higher abundances, with only $\sim24$~\% with lower values. Moreover, only five stars show \Nab{} values inconsistent with CAS within uncertainties, all of them SB1.
Overall, \Nab{}\,$\approx\!8.45$~dex emerges as an empirical upper limit for nitrogen enrichment without accompanying helium enrichment during the MS phase for dwarfs and giants.

The remaining 19 stars ($\sim$16\% of the analyzed sample) are He-rich, with an overall SB1 incidence of $\sim$33\%. The N abundance distribution of this latter group is bimodal, with a first peak around \Nab{}\,$\sim\!8.1$~dex and a second around \Nab{}\,$\sim\!8.5$~dex, separated by a gap at \Nab{}\,$\sim\!8.4$~dex. Among the tail of stars with lower N abundances (11 stars, none of them with an abundance below the CAS value), the SB1 fraction is $\sim$40\%, while in the higher-\Nab{} group (eight stars) it decreases to $\sim$33\%. Five out of the 9 stars comprising the high N abundance peak are classified as ON (a further discussion on the others is presented in Appendix~\ref{sec_ON}).

\subsection{Low nitrogen sample}\label{sec_lowN}

Within the current theoretical framework, we cannot explain N abundances under birth values, as nitrogen is expected to be enhanced during the main sequence. Therefore, we assess the reliability of the N abundance estimates for the 10 stars (75\% SB1) with \Nab{} values below the CAS\footnote{Only those inconsistent with the CAS even within uncertainties.}. \cite{Simon-Diaz+26} found a similar result for stars with helium under CAS values (Group \#2 in their work). Similar to them, we test two possible explanations: (i) an overestimated microturbulence, or (ii) contamination of the continuum by an external source. 

We test the first hypothesis in App.~\ref{sect_micro_effect}, where we show that this subsample cannot be explained by microturbulence effects.
Alternatively, we examined the potential impact of the presence of light-pollution by a companion star on the continuum normalization of the analyzed spectrum in App.~\ref{app_dilution}. Our analysis shows that such dilution effect can cause spurious determinations of N abundances which might be lowered by $\sim$0.15~dex or even more in the case of undetected SB2 systems. This would be coherent with the high binarity incidence in this group (text and third column in Table~\ref{tab_He_mic}).
Notably, the same effect similarly influences the derived \He{} abundances, consistent with the higher frequency of N-low stars among the He-low subsample --both effects likely share a common origin. 

\subsection{Distribution of stars in the Hunter diagram: general considerations}

\begin{figure*}[!t]
\centering
\includegraphics[width=.93\hsize,trim={0 1 0 0}]{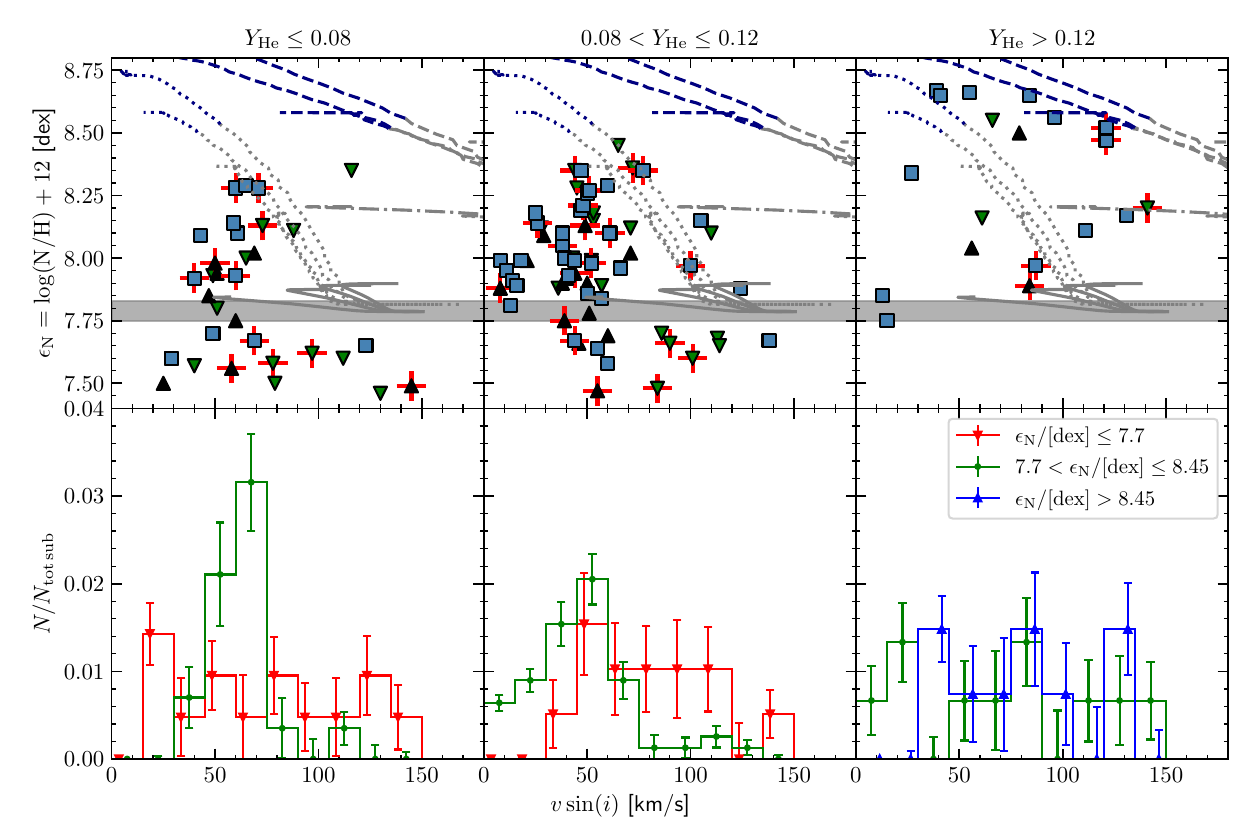}
  \caption{\textit{Top:} Surface N abundance versus projected rotational velocity. From left to right, He-low, normal, and rich sample. For reference pupposes, we include evolutionary tracks for $v_{\rm ini}/v_{\rm crit}=0.2$ and 0.4 from GENEC \citep[][and priv. comm.; dotted and dashed lines; $M_{\rm ini}=$\,20, 25, 32, and 40 $M_{\odot}$]{Ekstrom+12} and MESA models adopting the Mix1 prescription from \cite{Keszthelyi+22} \citep[equivalent to][solid and dashed dotted lines: $M_{\rm ini}=$\,20, 25, 30, and 40 $M_{\odot}$]{Brott+11} . Model segments with \He{}$\geq0.12$ are highlighted in blue. Same marker used as in Fig.~\ref{fig_sHR}.
  \textit{Bottom:} For the subsets in the upper panels, \vsini{} distributions separated in low, normal and high nitrogen (red, green and blue histograms, respectively).\\ 
 \label{fig_Hunter}}
\end{figure*}

Although our study focuses primarily on the regime of slow rotators, we can still draw meaningful conclusions regarding the role of rotational mixing in the chemical enrichment of the stellar surface.
To this end, we present in the upper row of Fig.~\ref{fig_Hunter} the so-called ''Hunter diagram" \citep[introduced by][]{Hunter+08}, displaying the relation between N abundance and \vsini{} for the sample. We separated the sample in the three groups according to their surface He abundance. This provides a complementary view to Figs.~\ref{fig_letter} and \ref{fig_histNitrogen_byHe}. From left to right, the panels correspond to the He-low, He-normal, and He-rich subsamples. The symbols indicate luminosity class (same code as in Fig.~\ref{fig_sHR}). 
We note that in this diagram we display the present day \vsini{}, which may differ significantly from the rotational velocity at birth due not only to projection effects, but to the braking during the evolution (depending mainly on the mass loss, magnetic field and angular momentum transport). In fact, the initial rotational velocity should be higher than the present day $v_{\rm rot}$ to enrich the surface with nitrogen in single stellar evolution.

For reference, we compare our observations with evolutionary tracks computed with the GENEC code \citep[][and private communication]{Ekstrom+12} for main-sequence stars with initial masses between 20 and 40~$M_{\odot}$ and two initial rotational velocities (0.2 and 0.4 $v_{\rm ini}/v_{\rm crit}$; dotted and dashed lines, respectively). We also include MESA models using the Mix1 prescription from \cite{Keszthelyi+22} \citep[equivalent to the models by][hereafter referred as "Mix1 MESA models"]{Brott+11} for the same initial masses, initial rotational velocities and initial chemical composition \citep[solid and dash-dotted lines, respectively; described in App.~A in ][]{Martinez-Sebastian+25}. We select these models as they include different prescriptions, particularly of convection (Schwarzschild criterion with an overshooting edge vs. Ledoux criterion with semi-convective area) and rotational mixing (transport of angular momentum and nuclides due to meridional circulation and turbulent shear vs. diffusive formalism including magnetic fields); both works consider the mass loss prescriptions from \cite{Vink+00,Vink+01}. Those represent two standard yet different assumptions in single stellar evolutionary models, leading to a more efficient mixing and surface braking in \cite{Ekstrom+12}.
The models are color-coded by He abundance: gray for \He{}\,$\leq\!0.12$ and blue for \He{}\,$>\!0.12$.
Under each panel, we included the \vsini{} distribution separated in three N abundance subgroups (\Nab\,$\leq\!7.7$~dex, $7.7\!<$\,\Nab\,$\leq\!8.45$~dex, and \Nab\,$>\!8.45$~dex; red, green, and blue histograms, respectively).

The two main N abundance groups identified in Fig.~\ref{fig_histNitrogen_byHe} among He-low stars (\He{}\,$\leq\!0.08$) are also evident in the left panel of Fig.~\ref{fig_Hunter}. Approximately 30\% of them correspond to low \Nab{} stars (discussed in Sect.~\ref{sec_lowN}) and have a fairly flat \vsini{} distribution. The remaining $\sim$70\% of these stars show normal to moderately enriched nitrogen (\Nab{}\,$<\!8.4$), and their \vsini\ distribution is concentrated in the range between $\sim$40 and 85~\kms. The distribution of stars from this latter group in the Hunter diagram is quite similar to the corresponding one in the He-normal group, but lacking stars in the low and high tails of the \vsini\ distribution.

The same two groups of N abundances are identified in stars with $0.08\!<$\,\He{}\,$\leq\!0.12$ (middle panels) but, as discussed in Sect.~\ref{sec_GeneralResults} (also Fig.~\ref{fig_histNitrogen_byHe}), the percentage of stars with abnormally low N abundances is much smaller. Again, the group of stars with normal to moderately enriched N abundances \citep[up to \Nab{}\,$\sim\!8.5$, group 2 in][]{Markova+18} are mostly concentrated at \vsini{}$\lesssim$60~\kms ($\sim$80\%); however, four of them are also found at larger \vsini{} values ($\geq$100~\kms{}). Similarly to the He-low group, the \vsini{} distribution is fairly flat for the N-low stars (\Nab{}\,$\leq\!7.7$~dex; red histogram) in the He-normal sample.

In the He-rich regime (\He{}\,$>\!0.12$; right panel), two clearly differentiated groups are observed, corresponding to the bimodal distribution peaks in Fig.~\ref{fig_histNitrogen_byHe}. The very N-rich group (upper part) comprises eight stars with \Nab{}\,$\sim\!8.6$~dex, five of them classified as ON (all the stars with this qualifier in the sample). They show an approximately flat distribution in \vsini{}, ranging from $\sim\!30$ to $\sim\!130$~\kms{} . The second group consists of nine stars centered around \Nab\,$\sim\!8.1$~dex, but with a larger internal scatter ($\Delta$\Nab\,$\sim\!0.25$~dex). These stars span the full range of projected rotational velocities covered in this study and show no evident correlation between \Nab{} and \vsini{}. Finally, two stars with very low projected rotational velocities exhibit CAS nitrogen abundances; however, their helium content (\He{}\,$=\!0.12$) places them at the boundary of the He-rich regime, making them also consistent with the He-normal population.

\section{Comparison with evolutionary models}\label{sec_discussion}

\subsection{Helium rich sample}\label{sec_high_He}  

Among the two sets of evolutionary models considered in Fig.~\ref{fig_Hunter}, only GENEC models with $v_{\rm ini}/v_{\rm crit}>0$ reproduce surface He abundances above \He{}\,$\geq\textbf{}\!0.12$ during the main sequence, and these cases are always accompanied by significant N enrichment (\Nab{}\,$\geq\!8.45$~dex; Fig.~\ref{fig_letter}). Consequently, all stars in our He-rich subsample (right panel of Fig.~\ref{fig_Hunter}) would be expected to display at least this level of nitrogen enrichment if they followed single-star evolutionary paths. However, several objects deviate from this prediction, showing enhanced helium but comparatively low N abundances. These inconsistent stars correspond to the lower-N peak in the He-rich distribution of Fig.~\ref{fig_histNitrogen_byHe} and to the lower region of the right-hand side of Fig.~\ref{fig_letter}, which we have argued to be binary products \citep[][ and Sect.~\ref{sec_GeneralResults}]{Martinez-Sebastian+25}.

In the framework of single stellar evolution, the eight He-rich stars that are also N-rich can only be reproduced by models assuming very high initial rotational velocities ($v_{\rm ini}/v_{\rm crit}\!>\!0.2$). Five of these stars are classified as ON, while the others are justified to be so in Appendix~\ref{sec_ON} and hereof will be referred as so.
They rotate, on average, at \vsini{}\,$\sim\!80$~\kms{}, compared to \vsini{}\,$\sim\!60$~\kms{} for the He-normal sample. Also, the associated \vsini\ distribution seems to be fairly flat. 
However, this can be a low number effect, highlighting again the need to increase the statistics for the He-rich group. 

\begin{figure}[!t]
\includegraphics[width=1.\hsize,trim={0 1 0 0}]{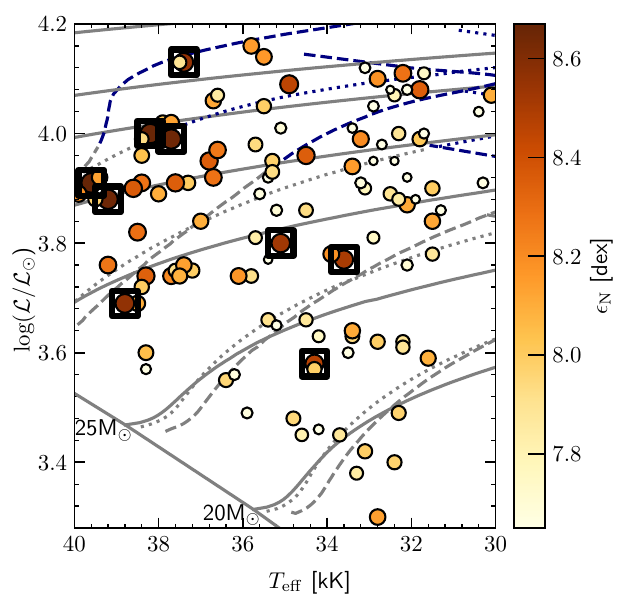}
  \caption{sHRD of the analyzed sample, with color and marker size scaled to N abundance. Solid black squares highlight stars with \He{}\,$>\!0.12$ and \Nab{}\,$>\!8.45$~dex. As a reference, GENEC tracks with $v_{\rm ini}/v_{\rm crit}\!=\!0$, 0.2, and 0.4 (solid, dotted and dashed lines, respectively.) Blue segments correspond to phases with \Nab{}\,$\geq$\,8.45~dex. \label{fig_sHR_N}}
\end{figure}

We highlight these eight targets with black-edged squares in Fig.~\ref{fig_sHR_N}, where the color (and marker size) represents the surface N abundance. As a reference, we include GENEC evolutionary tracks with $v_{\rm ini}/v_{\rm crit}\!=\!0$, 0.2, and 0.4 (solid, dotted, and dashed lines, respectively), marking in blue the regions where \Nab{}\,$\geq\!8.45$. The stars under discussion appear more enriched than their neighbors, yet they do not occupy any particular region of the sHRD. Moreover, when compared with predictions from \cite{Ekstrom+12} models with high initial rotational velocities, five of them appear too close to the ZAMS to be He enriched through single stellar evolution.
This is consistent with previous findings \citep[e.g.][]{Martins+15B,Simon-Diaz+26}, which showed that ON stars cannot be systematically reproduced by single-star evolutionary models.
\cite{Rivero-Gonzalez+12} found a similar N and He surface overenrichment in the bulk of O-type stars in the LMC, and interpreted it as an indication of efficient mixing during the very early phases of stellar evolution of these stars.

Alternatively, \cite{Bolton&Rogers78} proposed a binary-interaction origin for ON stars, which subsequent observational studies have reinforced \citep[e.g.][]{Boyajian+05,Li&Howarth20}. In this framework, mass transfer can reproduce the observed surface abundances \citep{Jin+26} while changing the \Teff{} and \logg{} of the star, mimicking the position of younger objects in the sHRD \citep{McCrea64}. This scenario also provides a straightforward explanation for the higher average rotational velocities observed among ON stars, as spin-up is an expected outcome of mass accretion \citep{deMink+13,Benvenuto+25}. Moreover, the fact that they are far from critical rotation would point to case A mass transfer preventing the gainers spinning up to critical velocities \citep[][]{Langer12,Simon-Diaz+26}.

The subsequent evolution of the binary system may result in a runaway star following the supernova explosion of the initially more massive component \citep{Blaauw1961,LeonardDuncan1988}, potentially leading to a higher fraction of runaways among ON stars. Using complementary information from \cite{Carretero-Castrillo+23,Carretero-Castrillo+25}, and private communication, we find 50~\% of runaways among the stars which are simultaneously He- and N-rich, compared to only 15~\% in the He-normal sample. Noteworthy, this process can also account for the three stars that could, in principle, be reproduced by single-star evolution (located in the upper left corner of Fig.~\ref{fig_sHR_N}). Binary interaction therefore emerges also as the most plausible scenario for the physical origin of He-rich, N-rich stars, despite a single stellar origin cannot be rejected for some of them. This physical interpretation for the groups comprising all ON in our sample, but beyond the morphological classification, supports the discussion in \cite{Simon-Diaz+26}.

To summarize, we have presented observational evidence challenging the single star origin of He-rich O-type stars under the current paradigm. However, these characteristics are naturally explained if they are binary interaction products. This agrees with findings in \cite{Simon-Diaz+26} regarding He abundances in the full IACOB sample of Galactic O-type stars (not limited in \vsini{}).
Consequently, we argue that binary interaction is the dominant channel for surface helium enrichment in O-type stars.
In line with this, \cite{Cazorla+17} found that approximately half of their sample of 40 fast-rotating Galactic massive stars could not be reproduced by single-star evolutionary models. In particular, they reported unexpectedly high He enrichment and the presence of $\sim20\%$ of stars with low nitrogen abundances. This behavior is consistent with the binary-interaction scenario proposed in this work. Nevertheless, this interpretation should be further tested through dedicated analyses of N abundances in larger samples of fast-rotating Galactic O-type stars, explicitly considering the potential role of binary interaction.

\subsection{Normal helium sample}\label{sec_normal_He}

We argue that the He-low sample is dominated by spurious abundance determinations and hinders a meaningful comparison with single stellar evolutionary models. In addition, the star comprising the He-rich sample are most plausibly binary products. Therefore, the He-normal stars (middle panel in Fig.~\ref{fig_Hunter}) constitute a more optimal subsample for tracing single-star evolutionary trends. The removal of the two aforementioned groups from this sample produces a cleaner set of data to confront with single star evolutionary models. Nonetheless, it can still contain binary interaction products lacking clear observational signatures \citep[e.g.][]{Jin+26}. Moreover, the presence of binaries, even preinteraction, can play an unclear yet important role in the internal transport through tides \citep[][]{Sciarini+26}.

Within the sample of He-normal stars, the objects with \Nab{}\,$\geq\!7.9$~dex remain unreproducible by models without efficient rotational mixing (i.e. Mix1 MESA models). Although these models can reach N abundances up to \Nab{}\,$\sim\!8.25$~dex for $v_{\rm ini}/v_{\rm crit}=0.4$, the bulk of our observed stars exhibit substantially lower projected rotational velocities, which would require a strong surface braking mechanism than observed \citep{Holgado+22}.

\begin{figure*}[!t]
\centering
\includegraphics[width=.95\hsize,trim={0 1 0 0}]{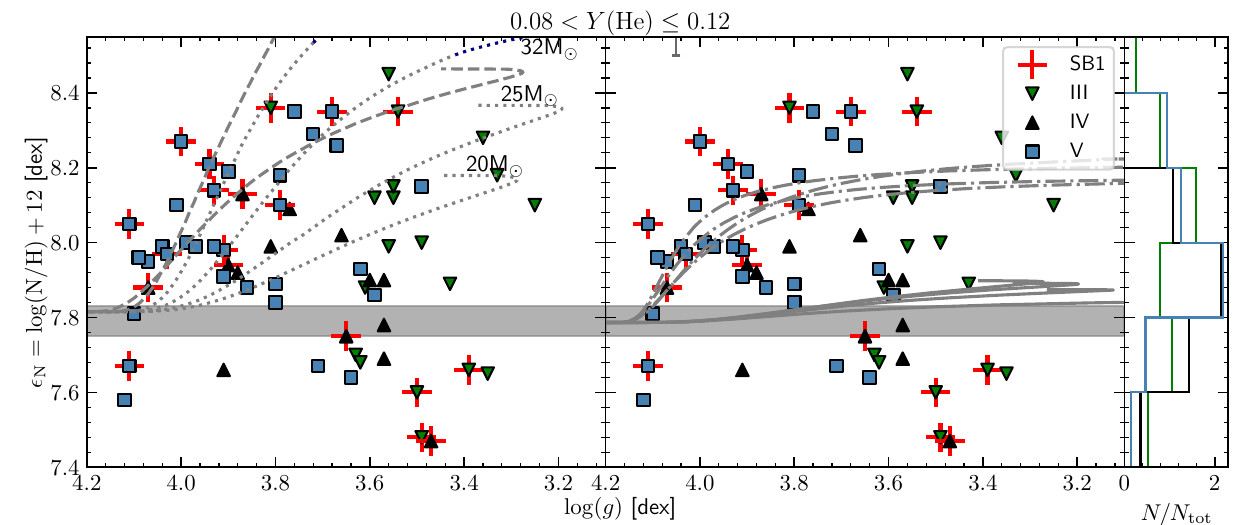}
  \caption{\textit{Left:} N abundance as a function of surface gravity for the He-normal subsample. The dotted and dashed lines correspond to GENEC evolutionary tracks with $v_{\rm ini}/v_{\rm crit}=0.2$ and 0.4, respectively (the latter only including tracks for $M_{\rm ini}=20~M_{\odot}$ and $M_{\rm ini}=40~M_{\odot}$). The blue segments indicate regions where models reach \He{}$\geq0.12$.
  \textit{Middle: }Same targets as left but using Mix1 MESA models with $v_{\rm ini}/v_{\rm crit}=0.2$ and 0.4 (solid and dotted dashed lines, respectively) as reference.
\textit{Right:} Density histogram of N abundance grouped by luminosity class, using the same color scheme as in the left panel.
 \label{fig_logg_epsN}}
\end{figure*}

Models including more efficient rotational mixing \citep{Ekstrom+12} predict higher N abundances than observed (\Nab{}\,$>\!8.4$~dex compared to the measured range $7.7~{\rm dex}\!<\,$\Nab{}\,$\leq\!8.4~{\rm dex}$) for the group of stars\footnote{Note that these stars could have been born with higher rotational velocities and subsequently spun down.} with \vsini{}\,$\lesssim\!50$~\kms{}. Considering projection effects, these stars could be rotating up to $\sim\!50\%$ faster on average, which would bring them closer --still slightly slower-- to the $v_{\rm ini}/v_{\rm crit}=0.2$ evolutionary tracks. We expect that models computed with lower initial rotational velocities but adopting the same mixing prescriptions as \cite{Ekstrom+12} would better reproduce the bulk of our observations.


To further investigate surface enrichment throughout stellar evolution, we use \logg{} as a proxy for evolutionary status in left and central panels of Fig.~\ref{fig_logg_epsN}.
We include evolutionary tracks with initial masses between 20~$M_{\odot}$ and 40~$M_{\odot}$ and $v_{\rm ini}/v_{\rm crit}=0.2$ and 0.4. In the left pannel, models from GENEC (only extreme cases for $v_{\rm ini}/v_{\rm crit}=0.4$ are included); in the middle panel; Mix1 MESA models. In both cases, we use the same coding as in Fig.~\ref{fig_Hunter}).
The right panel presents the corresponding \Nab{} histogram, separated by luminosity class using the same color scheme.

Stars labeled as a LC V and IV occupy a similar gravity range (\logg{}\,$\sim\!4.15$–3.5~dex) and exhibit comparable abundance distributions, both peaking around \Nab{}\,$\sim\!7.9$~dex, consistent with GENEC evolutionary tracks with $v_{\rm ini}/v_{\rm crit}\!=\!0.2$.
LC~III stars occupy lower gravities (\logg{}\,$\sim\!3.6$–$3.3$~dex, with one outlier at \logg{}\,$\sim\!3.8$~dex) and are slightly more N enriched than LC IV and V, with their distribution peaking at \Nab{}\,$\sim\!8.1$~dex and reaching values up to \Nab{}\,$\sim\!8.4$~dex. However, their enrichment level is not as pronounced relative to dwarfs as expected from GENEC evolutionary models with the considered initial rotations. In line with the previous discussion, we expect that lower initial rotational velocities better reproduce the stars with modest enrichment and low \vsini{}. Nonetheless, given the initial rotation distribution reported by \cite{Holgado+22}, we would expect our sample to be dominated by stars born with $v_{\rm ini}/v_{\rm crit}\sim0.2$. If this were the case, and the stars in our sample are the result of subsequent braking, their present-day surface abundances would show lower enrichment than predicted by these models.

In this projection, the data for LC~III stars seem to be better reproduced by the Mix1 MESA models with $0.2\!\lesssim\!v_{\rm ini}/v_{\rm crit}\!\lesssim\!0.4$ (models in the middle panel of Fig.~\ref{fig_logg_epsN}). However, this is difficult to reconcile with the low projected rotational velocities observed in Fig.~\ref{fig_Hunter}.
Despite we expect models similar to \cite{Ekstrom+12} with lower initial rotational velocities to better reproduce LC~III stars in both projections, they would face difficulties reproducing at the same time LC~V objects with the largest \logg{}. Therefore, the difficulties faced by both of the analyzed set of models --two extreme cases of enrichment-- to reproduce the observed distribution of \Nab{}, indicates that rotational mixing alone is not enough to account for N enrichment in this sample.

\section{Conclusions}\label{sec_conclusions}

In this paper we present results from the N abundance analysis of a curated sample of 117 Galactic O-type stars with \vsini{}\,$\leq\!150$~\kms{} from the IACOB project. 
The sample has been selected to maximize the reliability of the N abundance analysis, ensuring homogeneity and quality results. This work provides a uniform and statistically significant dataset, allowing a robust discussion of N enrichment in slow rotating O-type stars.

From the total of 117 analyzed stars, 28 are in common with previous studies that also derived N abundances (Section~\ref{sec_bib_comparison}). We find an overall good agreement with them, with a small systematic offset toward lower abundances in this work (\Nab\,=\,$0.06\pm0.11$~dex), still consistent within the typical uncertainty ($\Delta$\Nab$\sim$0.15~dex). This comparison highlights the strong consistency of results when comparing large samples. 

We find distinct N abundance distributions across the three helium regimes defined in \cite{Simon-Diaz+26} (Sect.~\ref{sec_GeneralResults}).
Stars with \He{}\,$\leq\!0.08$ exhibit a higher fraction of objects with abnormally low N abundances compared to the rest of the sample (Sect.~\ref{sec_lowN}).
We argue that the origin of both the low N and He abundances is the same: spurious determinations caused by light pollution of the analyzed spectra by a close-by object.

On the opposite side, stars defined as He-rich (\He{}\,$>\!0.12$) display systematically higher N abundances and projected rotational velocities than the rest of the sample (Section~\ref{sec_high_He}). Within this group, we identify two distinct subpopulations based on their N abundance, none of them totally compatible with single stellar evolution.
The first consists of moderately N-rich stars, whose properties are consistent with the outcomes of past mass-transfer events \citep{Martinez-Sebastian+25}.
The second comprises a much more enriched group (\Nab{}\,$\geq\!8.45$~dex), whose locations in the sHRD are inconsistent with high He abundance predicted by single-star evolution even in the most mixing efficient models \citep[i.e.][]{Ekstrom+12}. We interpret them as mass gainers mimicking younger objects after mass accretion.

Given that both He-rich subgroups in our sample --highly or comparatively low N enriched-- are better explained as binary products, we propose that binary interaction is the dominant mechanism driving surface helium enrichment in slow rotating O-type stars. The faster mean rotation yet far from critical in the He-rich sample points to a case A mass transfer that prevents critical rotation. 
In this framework, helium overabundance in main-sequence massive stars could serve as a relatively simple yet powerful indicator of past binary interaction. A complementary study in this line is presented in \cite{Simon-Diaz+26}.

As a corollary of the previous discussions, the He-normal stars ($0.08\!<$\,\He{}\,$\leq\!0.12$) emerge as the most suitable -- yet contaminated -- subsample for investigating single-star evolution (Section~\ref{sec_normal_He}). Within this group, the N abundances of most LC IV and V are consistent with stars born as slow rotators ($v_{\rm ini}/v_{\rm crit}\!\lesssim\!0.2$) and evolving as single stars with efficient mixing. For giants, however, we do not observe the expected abundance pattern (i.e. systematically higher \Nab{}), although their distribution is shifted toward higher nitrogen values. In their case, models with less efficient mixing and higher initial rotation rates could better reproduce the observed surface abundances, but would fail to match the low projected rotational velocities. Therefore, we claim that rotational mixing alone is not enough to explain the N enrichment pattern in this group.

In conclusion, this study \citep[together with][]{Simon-Diaz+26} provides observational evidence supporting that the helium enrichment observed in Main Sequence massive stars is better explained in the current paradigm as a consequence of past mass-transfer events and constraints in models of single star evolution should not be addressed from them. In addition to He-abundances, the analysis of N and, in the future, C and O abundances for large samples of O-type stars is essential to draw further constraints in the internal mixing during the MS.

The selection criteria imposed on our sample in terms of effective temperatures and wind strengths helped to minimize potential biases. However, they also make further work including the so-far excluded stars necessary in order to investigate the late main-sequence phase of the most massive stellar objects. In the same vein, the methodology followed imposed a limit to only study slow rotators. A dedicated abundance study focused on fast rotators \citep[which have been proposed to be binary products;][]{deMink+13,Britavskiy+23} could provide additional evidence to test our hypothesis.


\begin{acknowledgements}
We thank the anonymous referee for their careful review of the manuscript and the consequent comments and suggestions provided.
C.M-S, G.H, and S.S-D acknowledges the support from the State Research Agency (AEI) Spanish Ministry of Science and Innovation (MICIN) and the European Regional Development Fund (FEDER) under grant PID2021-122397NB-C21 and grant PRODUCTOS DE LA INTERACCION DE ESTRELLAS MASIVAS REVELADOS POR GRANDES SONDEOS ESPECTROSCOPICOS,  with references PID2024-159329NB-C21. This publication made use of the IAC HTCondor facility (http://research.cs.wisc.edu/htcondor/), partly financed by the Ministry of Economy and Competitiveness with FEDER funds, code IACA13-3E-2493. The project leading to this application has received funding from European Commission (EC) under Project OCEANS - Overcoming challenges in the evolution and nature of massive stars, HORIZON-MSCA-2023-SE-01, No G.A 101183150 Funded by the European Union. C.M.S received funding from the ERASMUS+ Program -- student mobility for traineeship from the University of La Laguna (course 2024-2025). G.H. received the support from the “La Caixa” Foundation (ID 100010434) under the fellowship code LCF/BQ/PI23/11970035.
\end{acknowledgements}

%
%

\vspace{-5mm}
\bibliographystyle{aa}
\bibliography{references}


\begin{appendix} 

\section{Table of resulting stellar parameters} 

Table~\ref{tab_results} summarize the parameters of the sample. The two last columns correspond to the results of this work. Spectral Classification from GOSSS, when possible.

\onecolumn
\begin{longtable}{ccccccccc}
\caption{Spectroscopic parameters for the stars in this work. \textbf{Binarity status:} \textbf{SB1:} Single-lined spectroscopic binary; \textbf{LS:} likely-single star;  \textbf{MD:} Unclassified binarity due to less than three spectra in the IACOB database.\label{tab_results}}\\
\hline \hline
Star & SpC & \vsini & \Teff & \logg & \He$\times$100 & Binary & \Nab & $\xi$\\
 & & [\kms{}] & [kK] & [dex] & & status & [dex] & [\kms{}] \\
\hline
\endhead
HD~326~329   &   O9.7~V   &   87   &   32.3\,$\pm$\,0.9   &   3.94\,$\pm$\,0.15   & 12.3\,$\pm$\,2.9 &   SB1   &   7.97\,$\pm$\,0.15   &   10 \\ 
HD~239~729   &   O9.7~V   &   100   &   32.4\,$\pm$\,0.8   &   4.03\,$\pm$\,0.17   & 9.8\,$\pm$\,2.2 &   SB1   &   7.97\,$\pm$\,0.15   &   5 \\ 
HD~101~070   &   O9.7~V   &   59   &   32.8\,$\pm$\,0.9   &   4.15\,$\pm$\,0.17   & 7.1\,$\pm$\,1.2 &   MD   &   8.14\,$\pm$\,0.15   &   5 \\ 
BD~-13~4930   &   O9.7~V   &   8   &   33.1\,$\pm$\,0.9   &   4.04\,$\pm$\,0.21   & 9.8\,$\pm$\,1.9 &   LS   &   7.99\,$\pm$\,0.15   &   10 \\
HD~36~512   &   O9.7~V   &   13   &   33.3\,$\pm$\,0.7   &   4.08\,$\pm$\,0.12   & 12.1\,$\pm$\,2.2 &   LS   &   7.76\,$\pm$\,0.15   &   11 \\
HD~38~666   &   O9.5~V   &   111   &   33.4\,$\pm$\,0.7   &   3.88\,$\pm$\,0.13   & 12.2\,$\pm$\,3.0 &   LS   &   8.11\,$\pm$\,0.15   &   10 \\ 
HD~206~183   &   O9.5~IV-V   &   8   &   33.7\,$\pm$\,0.6   &   4.07\,$\pm$\,0.06   & 9.8\,$\pm$\,1.7 &   SB1   &   7.84\,$\pm$\,0.15   &   15 \\ 
HD~305~536   &   O9.5~V   &   60   &   34.2\,$\pm$\,1.0   &   4.12\,$\pm$\,0.19   & 8.9\,$\pm$\,2.0 &   MD   &   7.58\,$\pm$\,0.15   &   5 \\ 
HD~34~078   &   O9.5~V   &   13   &   34.6\,$\pm$\,0.6   &   4.10\,$\pm$\,0.11   & 9.9\,$\pm$\,2.2 &   LS   &   7.81\,$\pm$\,0.15   &   3 \\ 
HD~95~275   &   O9.2~V   &   25   &   33.9\,$\pm$\,0.4   &   3.79\,$\pm$\,0.10   & 8.6\,$\pm$\,1.3 &   MD   &   8.18\,$\pm$\,0.15   &   9 \\ 
HD~44~597   &   O9.2~V   &   15   &   34.2\,$\pm$\,1.1   &   3.91\,$\pm$\,0.20   & 12.1\,$\pm$\,4.1 &   LS   &   7.75\,$\pm$\,0.15   &   9 \\ 
HD~12~323   &   ON9.2V   &   121   &   34.3\,$\pm$\,0.9   &   3.93\,$\pm$\,0.19   & 18.2\,$\pm$\,5.1 &   SB1   &   8.47\,$\pm$\,0.15   &   12 \\ 
HD~58~465~A   &   O9.2~V   &   44   &   34.3\,$\pm$\,1.4   &   3.97\,$\pm$\,0.28   & 9.0\,$\pm$\,3.4 &   MD   &   7.99\,$\pm$\,0.22   &   19 \\ 
HD~46~202   &   O9.2~V   &   11   &   34.8\,$\pm$\,0.7   &   4.07\,$\pm$\,0.15   & 9.3\,$\pm$\,1.8 &   LS   &   7.95\,$\pm$\,0.15   &   6 \\ 
HD~306~097   &   O9~V   &   123   &   32.9\,$\pm$\,0.5   &   3.49\,$\pm$\,0.05   & 7.5\,$\pm$\,1.7 &   MD   &   7.65\,$\pm$\,0.23   &   10 \\ 
CPD~-59~2551   &   O9~V   &   124   &   34.5\,$\pm$\,0.6   &   3.86\,$\pm$\,0.07   & 11.6\,$\pm$\,1.8 &   LS   &   7.88\,$\pm$\,0.15   &   6 \\ 
HD~66~788   &   O9~V   &   29   &   35.2\,$\pm$\,0.8   &   3.93\,$\pm$\,0.12   & 6.9\,$\pm$\,1.4 &   MD   &   7.60\,$\pm$\,0.15   &   20 \\ 
HD~214~680   &   O9~V   &   14   &   35.4\,$\pm$\,0.6   &   3.91\,$\pm$\,0.08   & 10.8\,$\pm$\,1.8 &   LS   &   7.91\,$\pm$\,0.15   &   13 \\ 
HD~216~898   &   O9~V   &   44   &   35.9\,$\pm$\,1.0   &   4.11\,$\pm$\,0.17   & 9.7\,$\pm$\,2.2 &   SB1   &   7.67\,$\pm$\,0.15   &   15 \\ 
HD~298~429   &   O8.5~V   &   105   &   33.4\,$\pm$\,0.8   &   3.49\,$\pm$\,0.13   & 10.4\,$\pm$\,2.6 &   MD   &   8.15\,$\pm$\,0.15   &   13 \\ 
HD~14~633   &   ON8.5~V   &   121   &   35.1\,$\pm$\,0.5   &   3.80\,$\pm$\,0.06   & 16.9\,$\pm$\,3.5 &   SB1   &   8.52\,$\pm$\,0.17   &   13 \\ 
HD~48~279   &   O8.5~VzNstrvar?   &   131   &   36.1\,$\pm$\,0.9   &   3.87\,$\pm$\,0.15   & 15.4\,$\pm$\,2.8 &   LS   &   8.17\,$\pm$\,0.18   &   15 \\ 
ALS~15~196   &   O8.5~V   &   66   &   36.4\,$\pm$\,0.8   &   4.09\,$\pm$\,0.14   & 11.3\,$\pm$\,2.7 &   MD   &   7.96\,$\pm$\,0.15   &   5 \\ 
HD~145~217   &   O8~V   &   49   &   35.2\,$\pm$\,1.1   &   3.73\,$\pm$\,0.18   & 7.9\,$\pm$\,2.4 &   LS   &   7.70\,$\pm$\,0.15   &   15 \\ 
HD~97~848   &   O8~V   &   41   &   35.3\,$\pm$\,0.5   &   3.62\,$\pm$\,0.06   & 9.9\,$\pm$\,2.0 &   MD   &   7.93\,$\pm$\,0.15   &   14 \\ 
HD~101~223   &   O8~V   &   55   &   35.4\,$\pm$\,0.6   &   3.64\,$\pm$\,0.06   & 8.8\,$\pm$\,1.5 &   LS   &   7.64\,$\pm$\,0.15   &   11 \\ 
HD~101~191   &   O8~V   &   138   &   35.6\,$\pm$\,1.1   &   3.71\,$\pm$\,0.16   & 10.5\,$\pm$\,3.6 &   LS   &   7.67\,$\pm$\,0.31   &   15 \\ 
HD~191~978   &   O8~V   &   57   &   35.7\,$\pm$\,0.8   &   3.80\,$\pm$\,0.10   & 8.9\,$\pm$\,1.7 &   LS   &   7.84\,$\pm$\,0.15   &   10 \\ 
HD~305~438   &   O8~Vz   &   18   &   37.2\,$\pm$\,0.5   &   3.93\,$\pm$\,0.05   & 9.5\,$\pm$\,1.5 &   MD   &   7.99\,$\pm$\,0.16   &   17 \\ 
HD~99~546   &   O7.5~V((f))Nstr   &   50   &   36.6\,$\pm$\,0.7   &   3.67\,$\pm$\,0.09   & 11.2\,$\pm$\,1.9 &   MD   &   8.26\,$\pm$\,0.15   &   14 \\ 
HD~344~777   &   O7.5~V   &   71   &   36.7\,$\pm$\,0.9   &   3.79\,$\pm$\,0.14   & 8.0\,$\pm$\,0.9 &   SB1   &   8.28\,$\pm$\,0.15   &   10 \\ 
HD~46~573   &   O7.5~V((f))   &   77   &   36.8\,$\pm$\,0.7   &   3.68\,$\pm$\,0.07   & 11.7\,$\pm$\,2.2 &   SB1   &   8.35\,$\pm$\,0.15   &   16 \\ 
HD~168~504   &   O7.5~V   &   61   &   37.0\,$\pm$\,1.1   &   3.79\,$\pm$\,0.11   & 9.3\,$\pm$\,2.0 &   SB1   &   8.10\,$\pm$\,0.15   &   7 \\ 
HD~44~811   &   O7.5~Vz   &   26   &   37.4\,$\pm$\,0.7   &   3.93\,$\pm$\,0.13   & 10.6\,$\pm$\,1.8 &   SB1   &   8.14\,$\pm$\,0.15   &   7 \\ 
HD~35~619   &   O7.5~V((f))   &   40   &   37.6\,$\pm$\,0.7   &   3.94\,$\pm$\,0.14   & 7.2\,$\pm$\,1.3 &   SB1   &   7.92\,$\pm$\,0.16   &   17 \\ 
HD~152~590   &   O7.5~Vz   &   48   &   37.7\,$\pm$\,0.7   &   3.94\,$\pm$\,0.11   & 9.1\,$\pm$\,1.7 &   SB1   &   8.21\,$\pm$\,0.15   &   8 \\ 
BD~+60~586   &   O7.5~Vz   &   38   &   38.3\,$\pm$\,0.7   &   4.11\,$\pm$\,0.13   & 8.9\,$\pm$\,1.6 &   SB1   &   8.05\,$\pm$\,0.15   &   13 \\ 
HD~164~492   &   O7.5~Vz   &   39   &   38.4\,$\pm$\,0.8   &   3.99\,$\pm$\,0.13   & 9.4\,$\pm$\,1.9 &   LS   &   8.00\,$\pm$\,0.15   &   6 \\ 
HD~93~222   &   O7~V((f))z   &   50   &   36.6\,$\pm$\,0.7   &   3.59\,$\pm$\,0.09   & 11.1\,$\pm$\,2.3 &   LS   &   7.86\,$\pm$\,0.20   &   15 \\ 
HD~47~839   &   O7~V((f))z   &   43   &   37.5\,$\pm$\,0.6   &   3.90\,$\pm$\,0.12   & 6.2\,$\pm$\,1.0 &   LS   &   8.09\,$\pm$\,0.16   &   15 \\ 
HD~227~245   &   O7~V(n)((f))z   &   47   &   37.6\,$\pm$\,0.9   &   3.76\,$\pm$\,0.10   & 8.7\,$\pm$\,1.7 &   LS   &   8.35\,$\pm$\,0.15   &   17 \\ 
HD~193~595   &   O7~V((f))   &   41   &   37.7\,$\pm$\,0.8   &   3.72\,$\pm$\,0.13   & 15.1\,$\pm$\,3.4 &   LS   &   8.65\,$\pm$\,0.15   &   6 \\ 
HD~90~273   &   ON7~V((f))   &   55   &   38.2\,$\pm$\,0.7   &   3.71\,$\pm$\,0.07   & 18.0\,$\pm$\,4.2 &   MD   &   8.66\,$\pm$\,0.15   &   7 \\ 
BD~+56~594   &   O7~Vz   &   27   &   38.3\,$\pm$\,0.8   &   3.99\,$\pm$\,0.14   & 14.0\,$\pm$\,3.0 &   LS   &   8.34\,$\pm$\,0.15   &   14 \\ 
HD~93~146   &   O7~V((f))   &   60   &   38.4\,$\pm$\,0.7   &   3.79\,$\pm$\,0.12   & 7.4\,$\pm$\,1.4 &   SB1   &   7.93\,$\pm$\,0.15   &   18 \\ 
BD~+62~2078   &   O7~V((f))z   &   38   &   38.5\,$\pm$\,1.0   &   4.01\,$\pm$\,0.16   & 9.6\,$\pm$\,2.6 &   LS   &   8.10\,$\pm$\,0.15   &   11 \\ 
HD~110~360   &   ON7~Vz   &   96   &   38.8\,$\pm$\,1.3   &   4.060\,$\pm$\,0.2   & 18.3\,$\pm$\,4.7 &   MD   &   8.56\,$\pm$\,0.15   &   15 \\ 
HD~91~824   &   O7~V((f))z   &   51   &   39.2\,$\pm$\,0.7   &   4.00\,$\pm$\,0.09   & 8.2\,$\pm$\,1.6 &   SB1   &   8.27\,$\pm$\,0.15   &   13 \\ 
BD~+62~424   &   O6.5~V(n)((f))   &   60   &   37.8\,$\pm$\,1.0   &   3.72\,$\pm$\,0.14   & 8.2\,$\pm$\,2.2 &   LS   &   8.29\,$\pm$\,0.15   &   10 \\ 
HD~227~018   &   O6.5~V((f))z   &   61   &   38.0\,$\pm$\,1.0   &   3.85\,$\pm$\,0.14   & 6.0\,$\pm$\,2.2 &   LS   &   8.10\,$\pm$\,0.15   &   5 \\ 
HD~242~935   &   O6.5~V((f))z   &   29   &   38.3\,$\pm$\,1.1   &   4.17\,$\pm$\,0.21   & 7.0\,$\pm$\,1.5 &   MD   &   7.60\,$\pm$\,0.15   &   7 \\ 
HD~91~572   &   O6.5~V((f))z   &   60   &   38.4\,$\pm$\,0.7   &   3.81\,$\pm$\,0.12   & 7.6\,$\pm$\,1.7 &   SB1   &   8.28\,$\pm$\,0.15   &   10 \\ 
HD~344~784   &   O6.5~V((f))z   &   65   &   38.5\,$\pm$\,0.9   &   3.92\,$\pm$\,0.15   & 7.6\,$\pm$\,1.7 &   MD   &   8.29\,$\pm$\,0.19   &   9 \\ 
HD~12~993   &   O6.5~V((f)) Nstr   &   84   &   39.2\,$\pm$\,1.0   &   3.88\,$\pm$\,0.16   & 17.2\,$\pm$\,5.3 &   LS   &   8.65\,$\pm$\,0.15   &   16 \\ 
HD~199~579   &   O6.5~V(n)((f))z   &   52   &   39.5\,$\pm$\,0.9   &   3.91\,$\pm$\,0.12   & 8.7\,$\pm$\,1.7 &   SB1   &   7.98\,$\pm$\,0.23   &   11 \\ 
CPD~-58~2611   &   O6~V((f))z   &   39   &   39.6\,$\pm$\,1.0   &   3.85\,$\pm$\,0.12   & 12.5\,$\pm$\,2.9 &   MD   &   8.67\,$\pm$\,0.15   &   9 \\ 
HD~303~311   &   O6~V((f))z   &   47   &   39.9\,$\pm$\,1.0   &   3.90\,$\pm$\,0.11   & 9.7\,$\pm$\,1.9 &   MD   &   8.19\,$\pm$\,0.15   &   17 \\ 
BD~+45~3216~A   &   O5~V((f))z   &   69   &   36.2\,$\pm$\,0.8   &   4.11\,$\pm$\,0.15   & 6.4\,$\pm$\,1.0 &   SB1   &   7.67\,$\pm$\,0.15   &   5 \\ 
\hline
HD~207~538   &   O9.7~IV   &   29   &   31.6\,$\pm$\,0.6   &   3.77\,$\pm$\,0.07   & 10.8\,$\pm$\,1.3 &   LS   &   8.09\,$\pm$\,0.15   &   8 \\ 
HD~190~427   &   O9.7~IV   &   25   &   31.9\,$\pm$\,1.1   &   3.53\,$\pm$\,0.16   & 7.5\,$\pm$\,1.7 &   LS   &   7.50\,$\pm$\,0.15   &   15 \\ 
HD~209~339   &   O9.7~IV   &   84   &   32.2\,$\pm$\,0.7   &   3.82\,$\pm$\,0.07   & 12.1\,$\pm$\,3.0 &   SB1   &   7.89\,$\pm$\,0.15   &   11 \\ 
HD~232~525   &   O9.7~IV   &   21   &   32.8\,$\pm$\,1.1   &   3.81\,$\pm$\,0.17   & 10.6\,$\pm$\,3.1 &   MD   &   7.99\,$\pm$\,0.15   &   9 \\ 
HD~193~117   &   O9.5~IV(n)   &   60   &   31.7\,$\pm$\,0.8   &   3.32\,$\pm$\,0.13   & 7.1\,$\pm$\,1.6 &   LS   &   7.75\,$\pm$\,0.17   &   12 \\ 
HD~164~019   &   O9.5~IVp   &   69   &   31.8\,$\pm$\,0.5   &   3.40\,$\pm$\,0.08   & 7.8\,$\pm$\,1.9 &   LS   &   8.02\,$\pm$\,0.16   &   11 \\ 
HD~166~546   &   O9.5~IV   &   38   &   32.5\,$\pm$\,0.7   &   3.60\,$\pm$\,0.13   & 8.6\,$\pm$\,1.6 &   LS   &   7.90\,$\pm$\,0.20   &   20 \\ 
HD~192~001   &   O9.5~IV   &   44   &   33.4\,$\pm$\,0.9   &   3.90\,$\pm$\,0.15   & 9.2\,$\pm$\,2.2 &   SB1   &   7.94\,$\pm$\,0.15   &   8 \\ 
HD~93~027   &   O9.5~IV   &   46   &   33.5\,$\pm$\,0.8   &   3.91\,$\pm$\,0.13   & 10.7\,$\pm$\,2.9 &   MD   &   7.66\,$\pm$\,0.17   &   12 \\ 
HD~190~991   &   O9.2~IV   &   47   &   32.3\,$\pm$\,1.3   &   3.51\,$\pm$\,0.21   & 6.8\,$\pm$\,1.8 &   LS   &   7.85\,$\pm$\,0.15   &   18 \\ 
HD~164~438   &   O9.2~IV   &   55   &   32.4\,$\pm$\,0.8   &   3.47\,$\pm$\,0.10   & 9.4\,$\pm$\,2.8 &   SB1   &   7.47\,$\pm$\,0.15   &   18 \\ 
HD~96~622   &   O9.2~IV   &   39   &   32.9\,$\pm$\,0.8   &   3.65\,$\pm$\,0.12   & 10.2\,$\pm$\,2.3 &   SB1   &   7.75\,$\pm$\,0.15   &   5 \\ 
HD~76~341   &   O9.2~IV   &   51   &   33.1\,$\pm$\,1.1   &   3.57\,$\pm$\,0.15   & 10.2\,$\pm$\,3.5 &   LS   &   7.78\,$\pm$\,0.20   &   13 \\ 
HD~201~345   &   ON9.2~IV   &   79   &   33.6\,$\pm$\,0.9   &   3.72\,$\pm$\,0.14   & 13.0\,$\pm$\,4.2 &   LS   &   8.50\,$\pm$\,0.21   &   14 \\ 
HD~113~659   &   O9~IV   &   58   &   33.4\,$\pm$\,0.7   &   3.49\,$\pm$\,0.12   & 7.4\,$\pm$\,1.4 &   SB1   &   7.56\,$\pm$\,0.16   &   19 \\ 
HD~73~882   &   O8.5~IV   &   145   &   35.4\,$\pm$\,0.8   &   3.82\,$\pm$\,0.17   & 7.2\,$\pm$\,1.3 &   SB1   &   7.49\,$\pm$\,0.15   &   13 \\ 
HD~46~966   &   O8.5~IV   &   40   &   35.8\,$\pm$\,0.4   &   3.88\,$\pm$\,0.05   & 9.5\,$\pm$\,1.5 &   LS   &   7.92\,$\pm$\,0.15   &   10 \\ 
HD~135~591   &   O8~IV((f))   &   60   &   35.1\,$\pm$\,0.5   &   3.57\,$\pm$\,0.09   & 9.1\,$\pm$\,1.6 &   LS   &   7.69\,$\pm$\,0.15   &   18 \\ 
HD~168~444   &   O8~IV   &   50   &   35.3\,$\pm$\,0.7   &   3.67\,$\pm$\,0.13   & 7.0\,$\pm$\,1.3 &   SB1   &   7.98\,$\pm$\,0.15   &   11 \\ 
HD~97~319   &   O7.5~IV((f))   &   51   &   35.7\,$\pm$\,0.7   &   3.61\,$\pm$\,0.12   & 7.1\,$\pm$\,1.3 &   MD   &   7.94\,$\pm$\,0.15   &   10 \\ 
HD~99~897   &   O6.5~IV((f))   &   50   &   37.5\,$\pm$\,0.6   &   3.57\,$\pm$\,0.07   & 11.1\,$\pm$\,1.8 &   MD   &   7.90\,$\pm$\,0.20   &   23 \\ 
HD~101~298   &   O6.5~IV((f))   &   71   &   37.9\,$\pm$\,0.5   &   3.66\,$\pm$\,0.09   & 9.5\,$\pm$\,2.0 &   LS   &   8.02\,$\pm$\,0.15   &   16 \\ 
HD~63~005   &   O6.5~IV   &   56   &   38.4\,$\pm$\,1.7   &   3.77\,$\pm$\,0.22   & 13.2\,$\pm$\,5.2 &   MD   &   8.04\,$\pm$\,0.15   &   20 \\
HD~101~190   &   O6~IV((f))   &   49   &   39.4\,$\pm$\,0.9   &   3.87\,$\pm$\,0.13   & 8.8\,$\pm$\,1.6 &   SB1   &   8.13\,$\pm$\,0.16   &   17 \\ 
\hline
HD~13~022   &   O9.7~III   &   110   &   30.1\,$\pm$\,0.7   &   3.25\,$\pm$\,0.12   & 11.1\,$\pm$\,4.4 &   LS   &   8.10\,$\pm$\,0.15   &   8 \\ 
HD~156~234   &   O9.7~III   &   90   &   30.3\,$\pm$\,0.8   &   3.39\,$\pm$\,0.13   & 8.2\,$\pm$\,2.5 &   SB1   &   7.66\,$\pm$\,0.15   &   20 \\ 
HD~154~643   &   O9.7~III   &   101   &   31.3\,$\pm$\,0.8   &   3.50\,$\pm$\,0.14   & 9.8\,$\pm$\,3.8 &   SB1   &   7.60\,$\pm$\,0.15   &   20 \\  
HD~118~198   &   O9.7~III   &   43   &   31.5\,$\pm$\,0.7   &   3.49\,$\pm$\,0.11   & 10.2\,$\pm$\,3.8 &   MD   &   8.00\,$\pm$\,0.29   &   7 \\ 
HD~55~879   &   O9.7~III   &   28   &   31.5\,$\pm$\,0.7   &   3.55\,$\pm$\,0.12   & 10.6\,$\pm$\,3.3 &   LS   &   8.12\,$\pm$\,0.20   &   11 \\
HD~112~784   &   O9.7~III   &   36   &   31.5\,$\pm$\,0.7   &   3.61\,$\pm$\,0.13   & 9.0\,$\pm$\,2.2 &   MD   &   7.88\,$\pm$\,0.15   &   11 \\ 
HD~189~957   &   O9.7~III   &   88   &   32.1\,$\pm$\,0.4   &   3.55\,$\pm$\,0.08   & 6.6\,$\pm$\,1.0 &   LS   &   8.11\,$\pm$\,0.15   &   7 \\ 
HD~150~475   &   O9.7~III   &   86   &   32.1\,$\pm$\,0.6   &   3.63\,$\pm$\,0.09   & 9.0\,$\pm$\,1.5 &   MD   &   7.70\,$\pm$\,0.15   &   14 \\ 
HD~113~606   &   O9.5~III   &   40   &   30.4\,$\pm$\,0.6   &   3.29\,$\pm$\,0.11   & 7.5\,$\pm$\,1.7 &   MD   &   7.57\,$\pm$\,0.15   &   25 \\ 
HD~190~429~B   &   O9.5~II-III   &   112   &   31.7\,$\pm$\,0.6   &   3.41\,$\pm$\,0.10  & 7.7\,$\pm$\,2.0 &   LS   &   7.60\,$\pm$\,0.15   &   19 \\ 
CPD~-35~2105   &   O9.2~III   &   79   &   32.1\,$\pm$\,0.5   &   3.48\,$\pm$\,0.07   & 8.0\,$\pm$\,2.0 &   MD   &   7.50\,$\pm$\,0.15   &   18 \\ 
HD~16~832   &   O9.2~III   &   45   &   32.2\,$\pm$\,0.8   &   3.36\,$\pm$\,0.13   & 8.5\,$\pm$\,2.2 &   LS   &   8.28\,$\pm$\,0.15   &   8 \\ 
HD~171~201   &   O9~III   &   116   &   31.8\,$\pm$\,0.7   &   3.28\,$\pm$\,0.07   & 7.9\,$\pm$\,1.7 &   MD   &   8.35\,$\pm$\,0.15   &   9 \\ 
HD~229~234   &   O9~III   &   97   &   32.1\,$\pm$\,0.9   &   3.40\,$\pm$\,0.10   & 6.0\,$\pm$\,3.1 &   SB1   &   7.62\,$\pm$\,0.15   &   10 \\ 
HD~305~523   &   O9~II-III   &   57   &   32.3\,$\pm$\,0.7   &   3.43\,$\pm$\,0.07   & 11.3\,$\pm$\,3.0 &   MD   &   7.89\,$\pm$\,0.15   &   13 \\ 
HD~215~806   &   O9~III   &   51   &   32.4\,$\pm$\,0.7   &   3.39\,$\pm$\,0.11   & 7.3\,$\pm$\,1.4 &   LS   &   7.80\,$\pm$\,0.15   &   9 \\ 
HD~60~369   &   O9~III   &   78   &   32.7\,$\pm$\,0.7   &   3.49\,$\pm$\,0.08   & 7.9\,$\pm$\,2.1 &   SB1   &   7.58\,$\pm$\,0.15   &   18 \\ 
HD~109~978   &   O9~III   &   53   &   32.8\,$\pm$\,1.1   &   3.33\,$\pm$\,0.16   & 10.6\,$\pm$\,3.0 &   MD   &   8.18\,$\pm$\,0.15   &   15 \\ 
HD~113~904~B   &   O9~III   &   84   &   32.9\,$\pm$\,0.4   &   3.49\,$\pm$\,0.04   & 9.5\,$\pm$\,1.5 &   SB1   &   7.48\,$\pm$\,0.17   &   17 \\ 
HD~105~627   &   O9~III   &   141   &   33.2\,$\pm$\,0.6   &   3.45\,$\pm$\,0.08   & 13.6\,$\pm$\,3.9 &   SB1   &   8.20\,$\pm$\,0.15   &   11 \\ 
HD~96~654   &   O9~III   &   113   &   33.2\,$\pm$\,0.8   &   3.62\,$\pm$\,0.10   & 8.7\,$\pm$\,2.1 &   MD   &   7.68\,$\pm$\,0.20   &   16 \\ 
HD~24~431   &   O9~III   &   49   &   34.5\,$\pm$\,0.7   &   3.67\,$\pm$\,0.10   & 7.9\,$\pm$\,2.1 &   LS   &   7.93\,$\pm$\,0.15   &   7 \\ 
HD~344~863   &   O8.5/9~III   &   130   &   32.5\,$\pm$\,0.6   &   3.33\,$\pm$\,0.08   & 7.4\,$\pm$\,1.6 &   LS   &   7.46\,$\pm$\,0.15   &   22 \\ 
HD~116~852   &   O8.5~II-III((f))   &   114   &   33.1\,$\pm$\,0.8   &   3.35\,$\pm$\,0.11   & 8.1\,$\pm$\,1.9 &   LS   &   7.65\,$\pm$\,0.15   &   21 \\ 
HD~218~195   &   O8.5~IIINstr   &   44   &   34.5\,$\pm$\,0.7   &   3.54\,$\pm$\,0.12   & 11.6\,$\pm$\,1.8 &   SB1   &   8.35\,$\pm$\,0.15   &   5 \\ 
HD~173~820   &   O8~III   &   65   &   34.9\,$\pm$\,1.4   &   3.56\,$\pm$\,0.20   & 10.6\,$\pm$\,3.5 &   MD   &   8.45\,$\pm$\,0.15   &   17 \\ 
HD~36~861   &   O8~III((f))   &   52   &   35.5\,$\pm$\,0.6   &   3.56\,$\pm$\,0.08   & 9.6\,$\pm$\,1.4 &   LS   &   7.99\,$\pm$\,0.15   &   7 \\ 
HD~186~980   &   O7.5~III((f))   &   61   &   35.5\,$\pm$\,0.6   &   3.49\,$\pm$\,0.11   & 12.6\,$\pm$\,2.7 &   LS   &   8.16\,$\pm$\,0.15   &   9 \\ 
HD~163~800   &   O7.5~III((f))   &   53   &   35.8\,$\pm$\,0.5   &   3.55\,$\pm$\,0.08   & 8.4\,$\pm$\,1.0 &   LS   &   8.15\,$\pm$\,0.15   &   10 \\ 
HD~167~659   &   O7~II-III(f)   &   71   &   36.7\,$\pm$\,0.7   &   3.59\,$\pm$\,0.08   & 9.2\,$\pm$\,1.4 &   LS   &   8.12\,$\pm$\,0.21   &   20 \\ 
HD~156~738   &   O6.5~III(f)   &   65   &   37.3\,$\pm$\,1.0   &   3.78\,$\pm$\,0.14   & 7.0\,$\pm$\,1.6 &   MD   &   8.00\,$\pm$\,0.15   &   25 \\ 
HD~190~864   &   O6.5~III(f)   &   66   &   37.4\,$\pm$\,0.8   &   3.55\,$\pm$\,0.09   & 14.4\,$\pm$\,2.3 &   LS   &   8.55\,$\pm$\,0.15   &   9 \\ 
HD~152~723   &   O6.5~III(f)   &   73   &   37.7\,$\pm$\,0.7   &   3.68\,$\pm$\,0.09   & 7.7\,$\pm$\,1.2 &   SB1   &   8.13\,$\pm$\,0.30   &   20 \\ 
HD~96~946   &   O6.5~III(f)   &   72   &   38.6\,$\pm$\,0.9   &   3.81\,$\pm$\,0.13   & 11.0\,$\pm$\,3.0 &   SB1   &   8.36\,$\pm$\,0.15   &   15 \\ 
\hline
\end{longtable}
\twocolumn

\section{Selection criteria}\label{app_selection} 

\begin{figure*}[!t]
\centering
\includegraphics[width=0.8\hsize,trim={0 1 0 0}]{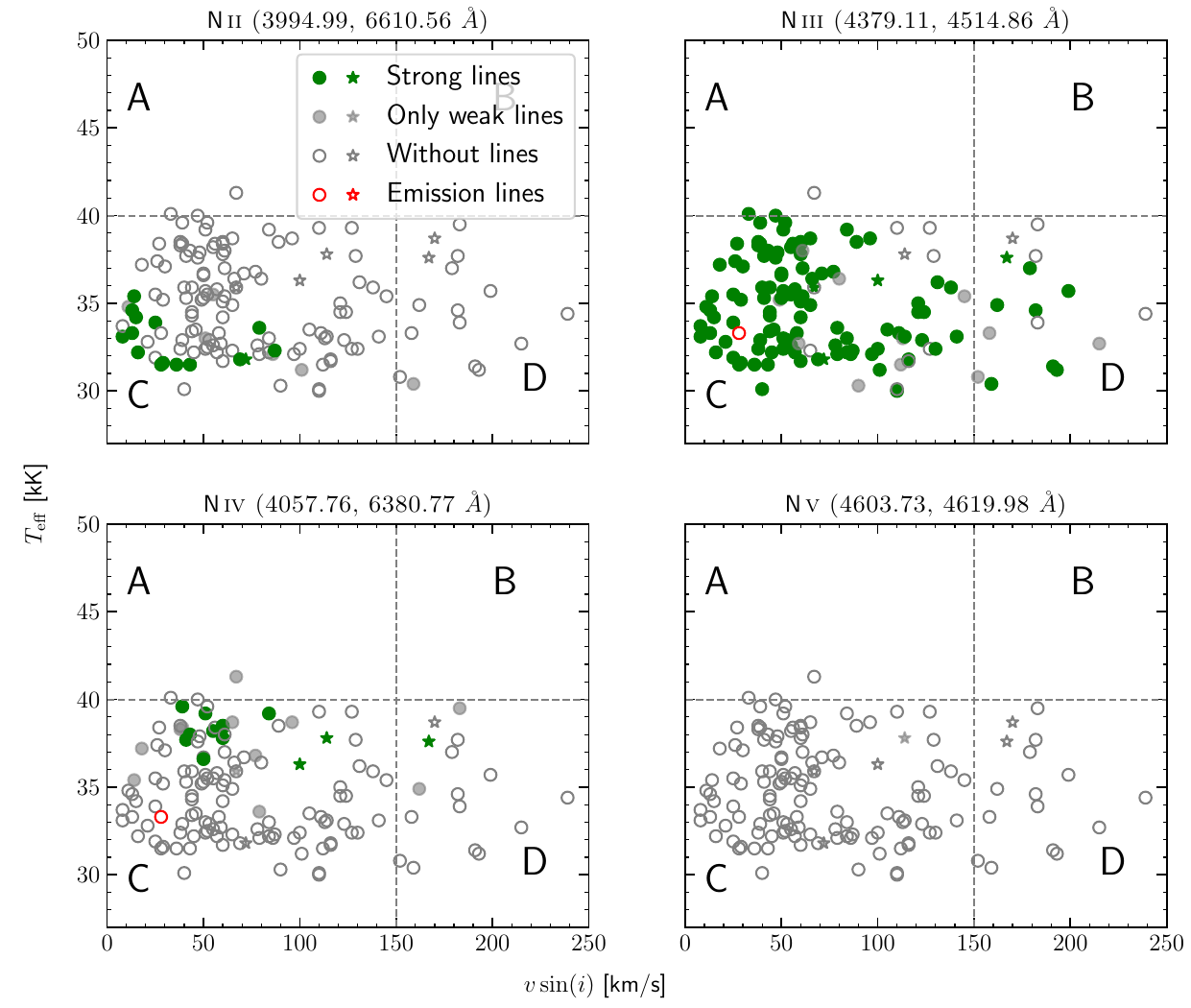}
  \caption{Summary of detection of key diagnostic lines from different N ions in the stars from the sample (Sect.~\ref{sec_sample}). From left to right, and top to bottom: detections of key diagnostic lines from \NII{}, \NIII{}, \NIV{}, and \NV{}, respectively. Filled black, gray, and empty gray symbols for stars displaying strong, weak and no lines, respectively. Circle and star symbols for stars with \logQ\,$\leq$\,12.7 and \logQ\,$>$12.7, respectively.
  Dotted lines indicate the final selection criteria; area marked as "C" corresponds to the final analyzed stars. 
 \label{fig_lines_presence}}
\end{figure*}

Figure \ref{fig_lines_presence} summarizes the results of the visual preanalysis described in Sect.~\ref{sec_sample}, shown in the \Teff{}\,--\,\vsini{} plane. As discussed there, we only include stars for which we have achieved a good fitting of the H and \HeI{}-{\sc ii} diagnostic lines when performing the IACOB-GBAT analysis \citep{Simon-Diaz+26}.
Black filled symbols indicate stars in which we clearly detected at least one of the indicated diagnostic lines. Gray filled symbols mark stars with detections of at least one line of the ion, although none of them strong. Open symbols represent stars where none of the visualized lines of the ion were found (Sect.~\ref{sec_sample}). Stars showing emission lines -- excluded from the analysis -- are marked in blue.
For completeness, we also include stars with winds stronger than \logQ{}\,$=\!-12.7$, indicated by star-shaped symbols and following the same color scheme. Only six such stars are present in the visualized sample of 145, proving that excluding them has a negligible effect on the overall sample statistics.

Our analysis reveals that, with the exception of two objects, all stars in the initially considered sample displaying nitrogen lines include \NIII{} among them. Due to the very limited number of \NIV{} and \NV{} lines, we restricted to stars also exhibiting \NIII{} lines (\Teff\,$\lesssim$\,40~kK).
For \vsini{}\,>\,150~\kms, nitrogen lines become so broadened that they are only clearly detected in only about half of the stars, whereas for lower \vsini{} values the detection rate exceeds 85\,\%. To avoid larger biases toward higher abundances, we limited the sample to slow rotators (\vsini\,$\leq$\,150~\kms). We note that a similar argument applies to stars with \vsini{}\,>\,100~\kms{}, as five targets with 100\,<\,\vsini\,<\,150~\kms\ show no detectable lines. However, as they represents only $\sim$15\% of this subsample, we decide to extend our analysis beyond \vsini{}\,=\,100~\kms{} in order to maximize our statistics while introducing only an acceptable bias.
The final analyzed sample corresponds to region~C in Fig.~\ref{fig_lines_presence}, since regions~A and~B contain very few stars, and the analysis of region~D would require a different approach (i.e. spectral synthesis instead of the curve-of-growth method).

\section{Methodological limitations}\label{app_errors}

In this appendix, we examine the limitations of our analysis and identify the dominant sources of uncertainty.

\subsection{Abundance determination consistency}\label{app_consistency}

First, we performed a consistency test to estimate the intrinsic uncertainty of our methodology. We generated a synthetic spectrum from a FASTWIND model and convolved it with instrumental, rotational, and macroturbulence profiles. The original model parameters appear in the second column of Table~\ref{tab_consistency_test}. We then added Gaussian noise centered at zero, with a standard deviation chosen to achieve SNR = 100.

We conduct a full blind analysis over this synthetic spectrum. To do so, we determine \vsini{} and \vmac{} with IACOB-BROAD \citep{Simon-Diaz&Herrero14}, the fundamental parameters with IACOB-GBAT \citep{Simon-Diaz+11} and the abundance following the methods from Sect.~\ref{sec_methods}. 
To estimate the effect the fundamental parameters determination itself, we repeat the abundance analysis with the original model parameters.

The results of both analyses, together with the original model parameters, are summarized in Table~\ref{tab_consistency_test}. The second column lists the original parameters used to generate the synthetic spectrum. The third column presents the parameters recovered in the blind analysis. The fourth column shows the results obtained when we fixed the stellar parameters (except for microturbulence and N abundance) to their original model values.

We find good agreement between the two analyses and the model values. In both cases, the derived abundances are consistent with the nominal value within the associated uncertainties. The differences between the nominal abundance and the recovered ones remain well below the typical minimum uncertainty adopted in our study.

The scatter in the abundance when fixing the parameters to the nominal values arises from the uncertainty in the $EW$ measurements. In contrast, in the blind analysis, we obtain a slightly higher abundance and a larger scatter, as expected from the propagation of uncertainties in the fundamental parameter determination. However, this effect on the final abundance remains smaller than the dispersion produced by the individual-line measurements.

\begin{table*}
\caption{Results of the analysis of a synthetic spectrum generated with a FASTWIND model.  The second column lists the parameters of the original model, the third column shows the results from the fully blind analysis, and the fourth column presents the abundance determination obtained when fixing the stellar parameters to the original model values. From top to bottom: effective temperature, gravity, wind strength, projected rotational, macroturbulence velocities, He abundance (by number) microturbulence velocity, and N abundance.}\label{tab_consistency_test}
\centering
\begin{tabular}{llll}
\hline \hline
Parameter & Model value & \multicolumn{2}{c}{Analysis result}\\
& & Blind analysis & Fixed parameters\\ 
\hline 
\Teff{} [kK] & 32.0 & $31.7\pm0.8$ & 32.0\\
\logg{} [dex] & 3.8 & $3.69\pm0.11$ & 3.8\\
\logQ{} & -13.5 & $-13.73\pm0.27$ & -13.5\\
\vsini{} [km/s] & 29 & 38 (using N\,II 3995) & 29\\
\vmac{} [km/s] & 50 & 54 (using N\,II 3995) & 50\\
\He{} &  0.10 & $0.10\pm0.02$ &  0.10\\
$\xi$ [km/s] & 11 & 8 & 12\\
\Nab{} [dex] & 8.1 & $8.15\pm0.11$ & $8.09\pm0.08$\\
\end{tabular}
\end{table*}

\begin{table}
\caption{Results of the abundance determination using different combinations of stellar parameters. In the table header, for reference, below the parameter in the column, the model value.
The fourth column shows the results obtained with the measured $EW$ values, while the fifth column presents the results when using the $EW$ values from the original model. We highlight with bold text the varied parameter.}\label{tab_errors}
\centering
\begin{tabular}{lllll}
\hline \hline
\Teff{} & \logg{} & $\xi$ & \Nab{}$_{;\,{\rm measured\,}EW}$ & \Nab{}$_{;\,{\rm nominal\,}EW}$ \\\
[kK] & [dex] & [\kms{}] & [dex] & [dex]\\
32.0 & 3.8 & 11 & 8.1 & 8.1 \\
 \hline
32.0 & 3.8 & 11 &  $8.10\pm0.08$   & $8.10\pm0.00$ \\
32.0 & 3.8 & \textbf{8}  &  $8.16\pm0.09$   & $8.14\pm0.03$  \\
32.0 & 3.8 & \textbf{14} &  $8.07\pm0.08$   & $8.08\pm0.02$  \\
\textbf{31.0} & 3.8 & 11 &  $8.06\pm0.18$   & $8.11\pm0.21$  \\
\textbf{33.0} & 3.8 & 11 &  $8.17\pm0.16$   & $8.13\pm0.21$  \\
32.0 & \textbf{3.7} & 11 &  $8.11\pm0.12$   & $8.08\pm0.12$  \\
32.0 & \textbf{3.9} & 11 &  $8.10\pm0.10$   & $8.12\pm0.08$  \\
\textbf{31.0} & \textbf{3.7} & 11 &  $8.06\pm0.13$   & $8.08\pm0.10$  \\
\textbf{33.0} & \textbf{3.9} & 11 &  $8.16\pm0.11$   & $8.13\pm0.13$  \\
\end{tabular}
\end{table}


We also investigated which is the dominant source of error in the abundance determination. To this end, we repeated the N abundance analysis on the same synthetic spectrum with different combinations of \Teff{}, \logg{}, and $\xi$ within typical uncertainties. In all runs, we set \vsini{}\,=\,29~\kms{}, \vmac{}\,=\,50~\kms{}, \He{}\,=\,0.10, and \logQ{}\,=\,$-13.5$.
To evaluate the impact of $EW$ uncertainties, we repeated each analysis using two sets of $EW$: (i) the measured values from the synthetic spectrum, and (ii) the $EW$ from the original FASTWIND model (the corresponding results are shown in the fourth and fifth columns of Table~\ref{tab_errors}, respectively). The first three columns of Table~\ref{tab_errors} list the parameters of the reference model used for the abundance determination. The uncertainty reported for each abundance corresponds to the scatter among the individual-line measurements.

Overall, we find a higher sensitivity to uncertainties in \Teff{} and $\xi$ than in \logg{}. However, the abundance variations introduced by these parameters remain smaller than the scatter produced by the individual-line measurements. This scatter therefore represents the dominant source of uncertainty in our analysis and covers those caused by the uncertainty in the fundamental parameters.

We note that the scatter increases significantly when the determined \Teff{} varies enough to affect the ionization balance. This situation is hardly rectifiable in observations where only one ionization stage is available, and it is mitigated when multiple stages are present, since we adjust \Teff{} to recover the correct ionization balance in nitrogen. In any case, this effect is expected to be smaller than the adopted minimum uncertainty of $\Delta$\Nab{}\,=\,0.15~dex. This value arises from typical uncertainties observed in stars for which several lines and ionization stages are available. Therefore, and as shown in this appendix, it supposes a conservative uncertainty.


\subsection{Impact of microturbulence in the estimation of helium abundances}\label{app_He}

\begin{figure}[!t]
\includegraphics[width=1.\hsize,trim={0 1 0 0}]{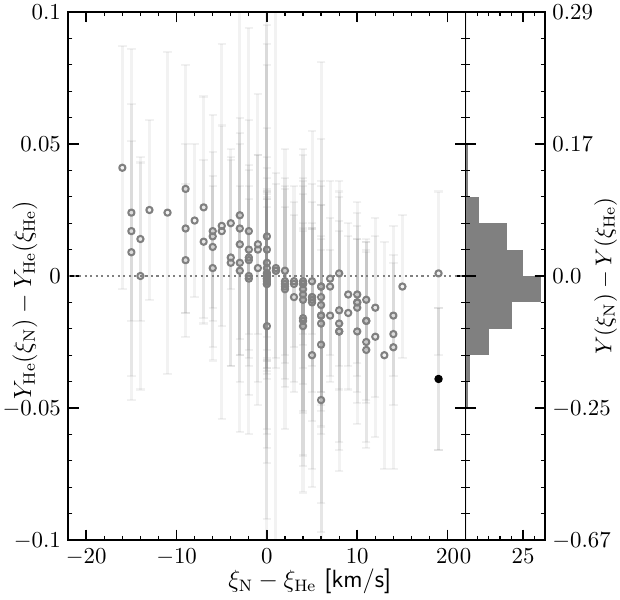}
  \caption{\textit{Left:} Difference in He abundance obtained using microturbulence derived from N and He abundance analysis, respectively, as a function of the difference between these microturbulence values. \textit{Right:} Histogram of the He abundance differences derived with the two $\xi$ determinations. \label{fig_He_mic_fixmic}}
\end{figure}

Despite its relevance in quantitative spectroscopy based on 1D model atmosphere calculations, the physical origin of microturbulence in massive stars remains poorly understood \citep{Markova+25}. As in previous works, we treat it as a free parameter and determine it simultaneously with the N abundance \citep[$\xi_{\rm N}$; e.g.][]{Carneiro+19,Rivero-Gonzalez+12}. A similar approach was adopted by \cite{Simon-Diaz+26} in their analysis of helium, deriving $\xi_{\rm He}$. However, as illustrated in Fig.~\ref{fig_He_mic_fixmic}, the resulting $\xi$ values can differ by up to $\sim15$~\kms{}. This difference might have a measurable impact in the estimated He abundances \citep[e.g.][]{McErlean98,Villamariz&Herrero00,Howarth&Smith01,Massey+13,Markova+20}, potentially affecting the conclusions of this work.
To assess this effect, we performed a consistency check by reanalyzing our sample with IACOB-GBAT \citep{Simon-Diaz+11,Sabin-Sanjulian+14,Holgado+18}, fixing the microturbulence to the values derived in this study with N lines.

\begin{figure}[!t]
\includegraphics[width=1.\hsize,trim={0 1 0 0}]{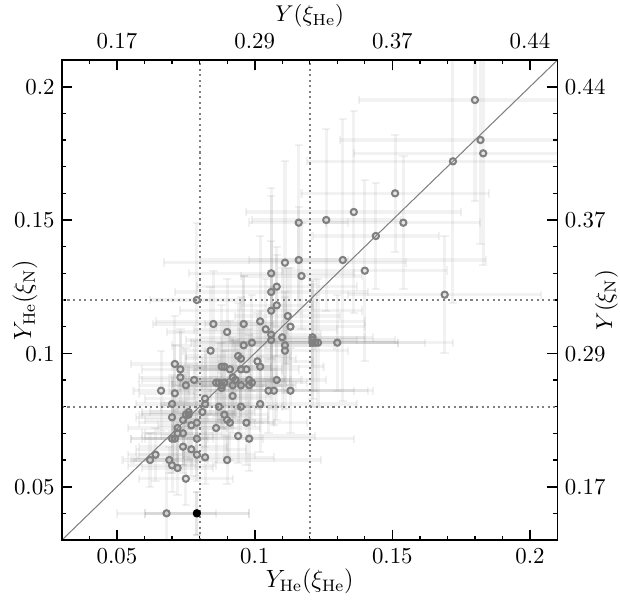}
  \caption{Comparison of the He abundances obtained using the microturbulence derived from N lines (this work) and from He analysis performed in \citep{Simon-Diaz+26}. \label{fig_He_fixmic}}
\end{figure}

In Fig.~\ref{fig_He_fixmic}, we compare the results of the reanalysis (\He{}$(\xi_{\rm N})$) with those from the reference works (\He{}$(\xi_{\rm He})$). As a reference, we include the one-to-one relation (solid diagonal line) and the limits defining He-low and He-rich regimes (dotted lines). Open symbols denote stars for which both determinations agree within the uncertainties. Notably, only one object shows a significant discrepancy (black point in the lower-left corner). In both studies, however, this star is classified as He-low.

Complementarily, in Fig.~\ref{fig_He_mic_fixmic} we show the difference in He abundance obtained with both microturbulence values ($\Delta Y_{\rm He}\!=\!Y_{\rm He}(\xi_{\rm N})\!-\!Y_{\rm He}(\xi_{\rm He})$) as a function of the difference in microturbulence ($\Delta \xi\!=\!\xi_{\rm N}\!-\!\xi_{\rm He}$); same marker coding as in Fig.~\ref{fig_He_fixmic}). The right panel presents the distribution of $\Delta$\He{} for the full sample. As expected, we find a clear correlation between both quantities. However, even for $|\Delta\xi|\sim\!15$~\kms{}, the resulting differences in He abundance remain comparable to the typical uncertainty ($\Delta$\He{}$\sim\!0.25$). Notably, the only star showing inconsistency between both determinations corresponds to the largest difference in $\xi$.

\begin{figure}[!t]
\includegraphics[width=1.\hsize,trim={0 1 0 0}]{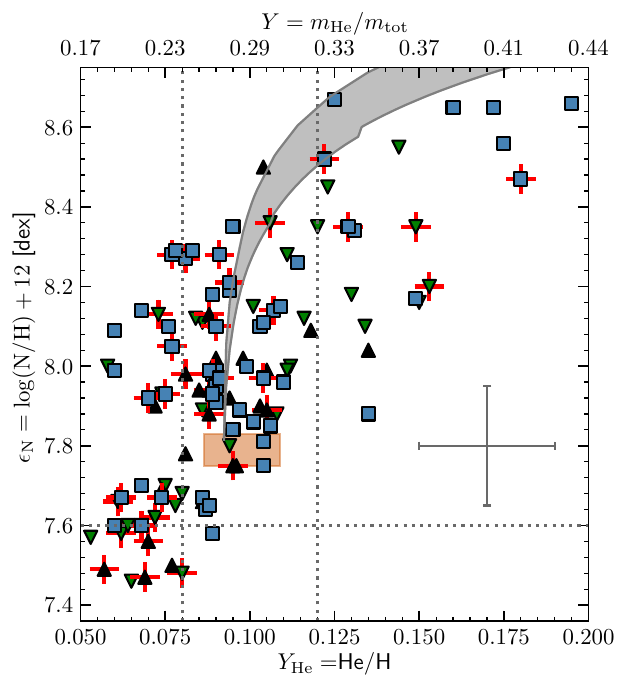}
\caption{Distribution of the sample in the N versus He abundance plane (equivalent to Fig.~\ref{fig_letter}) for He abundances derived with $\xi_{\rm N}$. \label{fig_letter_fixmic}}
\end{figure}

\begin{table}
    \centering
    \vspace{-0mm}
    \begin{tabular}{lllll}
        \hline\hline
     & \multicolumn{2}{c}{ \He{}$(\xi_{\rm He})$} & \multicolumn{2}{c}{\He{}$(\xi_{\rm N})$}\\
     \hline
     & Total & SB1 & Total & SB1 \\
     & subsample & [$\%$] & subsample & [$\%$]\\
    \hline
    He-low & 33 & 47.8 & 36 & 57.7\\
    He-normal & 65 & 42.2 & 62 & 33.3 \\
    He-rich & 19 & 33.3 & 19 & 41.7 \\
    \hline
    \end{tabular}
    \caption{Summary of the number of stars and the incidence of SB1 systems in the different He groups, constructed using He abundances derived with the two microturbulence determinations (from helium and nitrogen lines, respectively).\label{tab_He_mic}}
    \vspace{-5mm}
\end{table}

As a general trend, we find that some stars located near the boundaries between helium groups may shift classification depending on the adopted $\xi$. However, no systematic trend emerges when using either set of microturbulence values. As a final check, we include Fig.~\ref{fig_letter_fixmic} (equivalent to Fig.~\ref{fig_letter}, but using He abundances derived with $\xi_{\rm N}$). The overall distribution remains consistent with that shown in Fig.~\ref{fig_letter}. To quantify possible differences, we compare the number of stars in each helium group, as well as the fraction of SB1 systems, using both $Y_{\rm He}(\xi_{\rm He})$ and $Y_{\rm He}(\xi_{\rm N})$ (Table~\ref{tab_He_mic}). We find that the differences between groups remain within the expected uncertainties and do not affect the conclusions of this work.

Given these consistency checks, and in the absence of a physical argument favoring one microturbulence determination over the other, we adopt the He abundances derived using the microturbulence values obtained from the parallel analyses \citep[i.e.][]{Simon-Diaz+26} due to the self consistency of the methodology.

\subsection{Impact of microturbulence on the estimation of nitrogen abundances}\label{sect_micro_effect}

\begin{figure}[!t]
\includegraphics[width=1.\hsize,trim={0 1 0 25},clip]{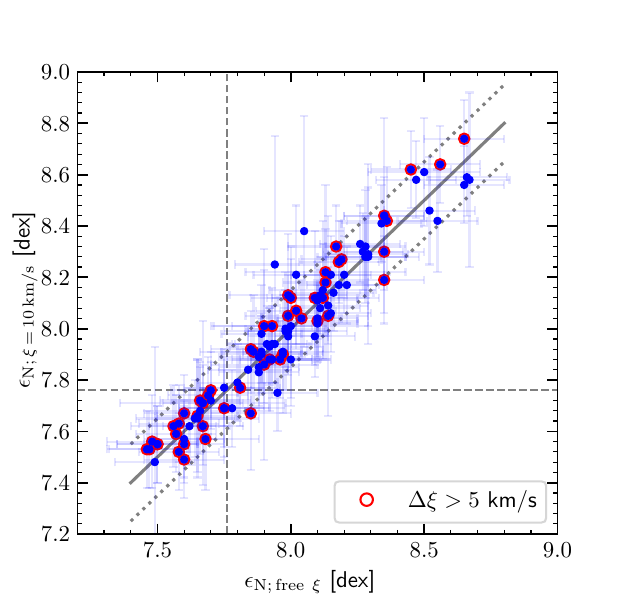}
  \caption{Comparison between N abundances derived fixing microturbulence to $\xi\!=\!10$~\kms{} and those obtained following the procedure in Sect.~\ref{sec_methods}. The solid line marks the 1-to-1 relation. The dotted lines indicate the typical uncertainty range. Stars with $\Delta\xi\!>\!5$~\kms{} are highlighted with red edges.
 \label{fig_abfree_ab10}}
\end{figure}

We have shown that the impact of a microturbulence uncertainty of $\sim$3~\kms{} is smaller than the typical abundance uncertainty (App.~\ref{app_consistency}). This effect may become relevant only when the microturbulence error is large or when it systematically biases the abundance. To test this possibilities, we repeated the analysis of the full sample by fixing $\xi\!=\!10$~\kms{}, a typical value for dwarfs.
Figure~\ref{fig_abfree_ab10} compares the abundances derived with fixed ($\xi = 10$~\kms{}) and free microturbulence. Dashed horizontal and vertical lines indicate the CAS values \citep{Nieva&Przybilla12}, while the solid and dotted diagonal lines mark the one-to-one relation and the expected agreement within the typical uncertainty ($\Delta$\Nab{}\,=\,0.15~dex), respectively. Stars with $\Delta\xi\!>\!5$~\kms{} are highlighted with red edges.

We find good agreement between both sets of results, with abundance differences smaller than the typical uncertainty for most stars. The six objects showing larger differences remain consistent once we consider their individual abundance uncertainties rather than the general value.

\subsection{Continuum dilution effect in He and N abundance determinations}\label{app_dilution}

In quantitative spectroscopy, additional light sources can significantly affect the analysis as they increase the continuum level, which results in the dilution of the spectral lines after normalization.
This effect may impact part of our sample. In this work, we classify as SB1 those systems that show no evidence of an additional component in the visual inspection of He~{\sc I}~5875 or the O~{\sc I}~7772–7775 triplet in any available spectrum. This criterion is consistent with previous studies in the IACOB series \citep[e.g.][]{Holgado+18,deBurgos+23}, but differs from the approach adopted in other recent works, which rely on spectral disentangling \citep[e.g.][]{Mayer+17,Mahy+22}. This is because we noted two main limitations of the latter method for our purposes: (i) disentangling is not available for the full sample, and (ii) the results can depend on the initial assumptions, leading to different classifications across studies (e.g. HD~163~892, for which \citeauthor{Barba+26} \citeyear{Barba+26} find no companion, while \citeauthor{Mahy+22} \citeyear{Mahy+22} classify it as an SB2). Since no disentangling has been performed for this work, we adopt the classical criterion based on visual inspection. In this context, some stars are known to host companions that contaminate the continuum of the primary (e.g. HD~91~824 or the eclipsing binary HD~152~590), while others may remain undetected.
Even in the absence of double-lined signatures, the companion may still contribute up to $\sim$10~\% of the continuum --larger flux ratios would typically reveal a secondary component in the spectra.
Additional contamination may also arise from unresolved nearby stars along the line of sight or from undetected SB2 systems, which could introduce even larger contributions to the continuum.

\begin{figure}[!t]
\includegraphics[width=1.\hsize,trim={0 1 0 0}]{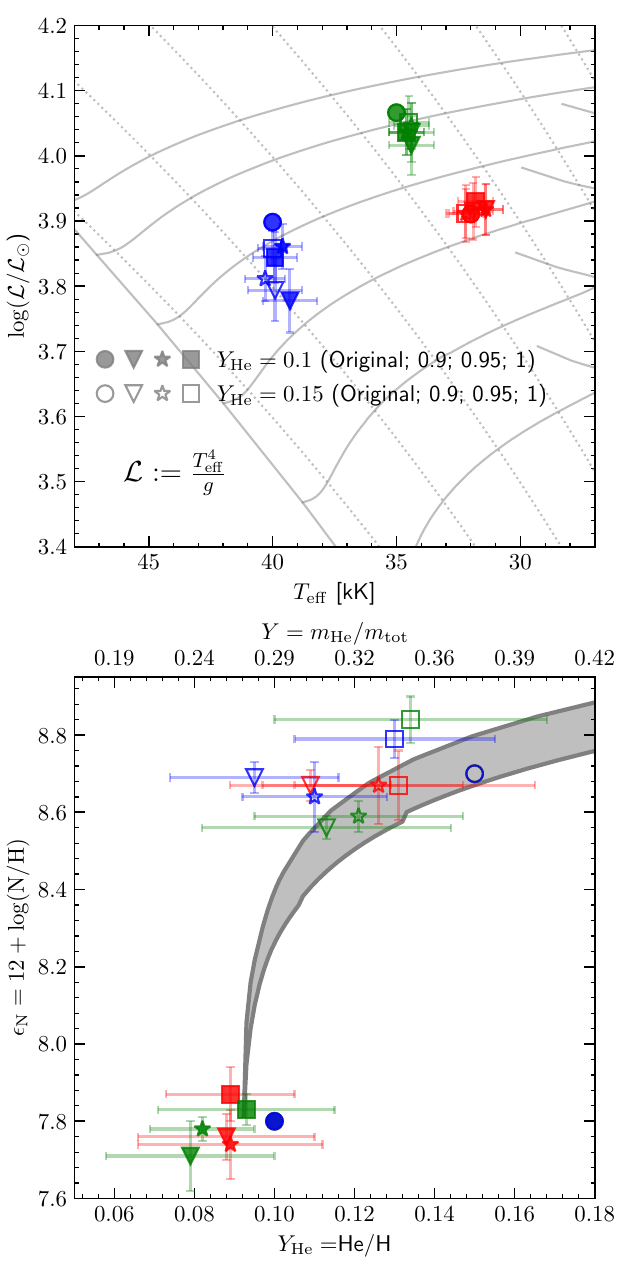}
  \caption{Results of the analysis of synthetic spectra from FASTWIND models. Six parameter combinations: \Teff{}–\logg{}\,=\,32~kK–3.5~dex, 35~kK–3.5~dex, and 40~kK–3.9~dex (red, green, and blue symbols, respectively), each with \He{}–\Nab{}\,=\,0.10–7.8~dex and 0.15–8.6~dex (filled and open symbols, respectively). For each model, we show the input values (circles) and the results obtained from the synthetic spectra with 0\%, 5\%, and 10\% continuum dilution (squares, stars, and triangles, respectively).
\textit{Top:} sHRD. For reference, we include the non-rotating evolutionary tracks from \cite{Ekstrom+12}. Dotted lines indicate constant gravities, starting at \logg{}\,=\,4.2~dex on the left and decreasing in steps of 0.2~dex.
\textit{Bottom:} N abundance versus He abundance (as in Fig.~\ref{fig_letter}). For reference, we include GENEC tracks for initial masses between 20 and 60~$M_{\odot}$ and $v_{\rm ini}/v_{\rm crit}\!=\!0.2-0.4$.\label{fig_dilution}}
\end{figure}

To analyze this effect, we followed a similar strategy to the one described in App.~\ref{app_consistency}. Namely, we used six FASTWIND models to generate synthetic spectra (circles in Fig.~\ref{fig_dilution}), covering three regions of the sHRD (\Teff{}–\logg{} combinations of 32~kK–3.5~dex, 35~kK–3.5~dex, and 40~kK–3.9~dex; red, green, and blue symbols, respectively) and two chemical compositions representative of ZAMS stars (\He{}\,=\,0.10 and \Nab{}\,=\,7.8~dex; filled symbols) and more evolved objects (\He{}\,=\,0.15 and \Nab{}\,=\,8.6~dex; open symbols).

For each case, we produced three synthetic spectra: one without dilution and two diluted to 95\% and 90\% of the original (square, star, and triangle symbols, respectively). We then analyzed these spectra —including the determination of the fundamental parameters— to quantify the impact of continuum dilution on the derived abundances.

The results are presented in Fig.~\ref{fig_dilution}. The top panel shows the sHRD with the derived fundamental parameters. For reference, we include the non-rotating evolutionary tracks from \cite{Ekstrom+12} (solid gray lines) and lines of constant surface gravity (gray dotted lines), starting at \logg{}\,=\,4.2~dex on the left and decreasing toward the right in steps of 0.2~dex. The bottom panel displays N versus He abundances, together with GENEC tracks for initial masses between 20 and 60~$M_{\odot}$ and $v_{\rm ini}/v_{\rm crit}\,=\,0.2-0.4$.

We do not find any significant effect of continuum dilution on the determination of \Teff{} from the He~{\sc i}–He~{\sc ii} ionization balance. In contrast, we detect an effect on \logg{}, which we associate with distortions in the wings of the Balmer lines. These distortions can mimic higher surface gravities. This effect is not uniform across the sHRD, becoming more pronounced at higher gravities.

For the models with \Teff{}\,=\,40~kK, \logg{}\,=\,3.9, \He{}\,=\,0.10, and \Nab{}\,=\,7.8~dex, we could not determine \Nab{} due to the weakness of the N lines, which prevents a reliable analysis. For the remaining models, we represented the uncertainty in \Nab{} as the scatter among the different analyzed lines, without imposing the minimum uncertainty of $\Delta$\Nab{}\,=\,0.15~dex —adopted elsewhere in this work.
In these cases, the continuum dilution leads to lower derived \Nab{} and \He{} values, mimicking the effect of lower surface abundances. 

As discussed above, this effect is expected to be more pronounced in undetected SB2 systems, where the contribution of the companion to the continuum is larger. To test this effect, we analyzed HD~93~028 and HD~101~413, two known SB2 systems composed of an O-type primary and an early B-type secondary, where the latter rotates significantly faster, making its detection more difficult \citep{Putkuri+26}.
For these stars, we derive \He{}\,=\,9.0\,$\pm$\,3.4 and \Nab\,=\,7.60\,$\pm$\,0.15, and \He{}\,=\,7.3\,$\pm$\,1.5 and \Nab{}\,=\,7.51\,$\pm$\,0.15, respectively. While these results are consistent with the expected dilution effect, the larger continuum contribution from the rapidly rotating companion leads to lower derived He and N abundances than in the synthetic model tests, where the dilution was limited to 10~\% (maximum).

It is important to note that none of the studied changes shift the derived abundances toward the region of He-rich and relatively low N. This rules out continuum dilution as a plausible explanation for the group of stars discussed in \cite{Martinez-Sebastian+25}.

\section{Revisiting the ON qualifier}\label{sec_ON}

A particularly relevant stellar group in the context of our study are the ON stars. This morphological classification correspond to O-type stars with enhanced N absorption, and C and O deficient compared to their spectral type and luminosity class \citep{Walborn76,Walborn03}. In particular, ON stars are delineated by the relative strength of \NIII{}~4634, \NIII{}~4640, and C\,{\sc iii}~4650 \citep{Sota+11}.
Their enhanced nitrogen lines may indicate N over-enrichment at the surface \citep{Martins+15B}. Table~\ref{tab_ON} summarizes the relevant information of the five objects of this type identified in our sample.
Interestingly, all of them are He-enriched, consistent with the results of \cite{Martins+15B} and \cite{Simon-Diaz+26}. They are also systematically N-enriched (\Nab{}\,$\geq\!8.45$~dex; most enriched bins in Fig.~\ref{fig_histNitrogen_byHe}); however, not all N-enriched objects are classified as ON. Since this classification is purely morphological, CNO anomalies should be interpreted in the context of the corresponding spectral type.

\defcitealias{Maiz-Apellaniz+16}{MA+16}
\defcitealias{Sota+11}{S+11}
\defcitealias{Sota+14}{S+14}
\begin{table*}
    \centering
    \begin{tabular}{llllllll}
        \hline\hline
          star & SpC & SpC reference & \Teff{} & \logg{} & $\log(L/L_{\odot})$ & \He{} & \Nab{} \\
           & & &[kK] & [dex] & & & [dex] \\
         \hline 
HD~90~273 & ON7\,V\,((f)) & \citetalias{Maiz-Apellaniz+16} & $38.2\pm0.7$ & $3.71\pm0.07$ & $5.13\pm0.08$ & $0.18\pm0.04$ & $8.66\pm0.07$\\
HD~110~360 & ON7\,V\,z & \citetalias{Maiz-Apellaniz+16} & $38.7\pm1.4$ & $4.03\pm0.20$ & $4.88\pm0.07$ & $0.19\pm0.05$ & $8.56\pm0.13$\\
HD~14~633 & ON8.5\,V & \citetalias{Sota+11} & $35.0\pm0.4$ & $3.79\pm0.07$ & $4.7\pm0.28$ & $0.18\pm0.03$ & $8.52\pm0.17$\\
HD~201~345 & ON9.2\,IV & \citetalias{Sota+14} & $33.6\pm0.9$ & $3.71\pm0.14$ & $4.81\pm0.18$ & $0.13\pm0.04$ & $8.5\pm0.21$\\
HD~12~323 & ON9.2\,V  & \citetalias{Sota+14} & $34.5\pm1.1$ & $3.95\pm0.20$ & $4.68\pm0.17$ & $0.18\pm0.07$ & $8.47\pm0.1$\\
\hline
        \hline
    \end{tabular}
    \caption{Stellar parameters and chemical abundances of the five ON stars in our sample. For each object: name, spectral classification (with reference), effective temperature, surface gravity, luminosity, and helium and nitrogen abundances.} 
    \label{tab_ON}
\end{table*} 

As a visual approach to place N-rich stars in an evolutionary context, Fig.~\ref{fig_sHR_N} shows the sample of stars in the sHRD, with color and symbol size indicating \Nab{}. As reference, we include GENEC evolutionary tracks for initial rotational velocities $v_{\rm ini}/v_{\rm crit}$\,=\,0, 0.2, and 0.4 (solid, dotted, and dashed lines, respectively), distinguishing between models with \Nab{}\,$<\!8.45$~dex (gray) and \Nab{}\,$\geq\!8.45$~dex (blue). Stars with \He{}\,$\geq\!0.12$ and \Nab{}\,$\geq\!8.45$~dex are highlighted with black squares.

All these stars display comparable abundances and are more enriched than other objects with similar spectral type, but not all of them are classified as ON. To investigate this, we revisited the morphological classification of those stars by directly comparing their spectra with those of reference ON stars of similar spectral type and luminosity class (Table~\ref{tab_ON_rev}) and revisiting the relative strength of  \NIII{}~4634, \NIII{}~4640 and C\,{\sc iii}~4650. Based on these comparisons, we propose stars in the firs column of Table~\ref{tab_ON_rev} to be reclassified as ON.

Our proposed additions are consistent with a correlation between line strengths and nitrogen enrichment, reinforcing the physical meaning of this morphological classification. Moreover, this exercise illustrates the diagnostic potential of quantitative analyses in spectral classification, as $EW$ remain unaffected by extrinsic line broadening.

\begin{table}
    \centering
    \begin{tabular}{llll}
        \hline\hline
          Studied & SpC & Reference & SpC \\
          star & & star &  \\    
         \hline
         
HD~190~864 & O6.5~III~(f) & HD~96~946 & O6.5~III~(f)\\
HD~193~595 & O7~V~((f)) & HD~90~273 & ON7~V~((f))\\
HD~12~993 & O6.5~V~((f))Nstr & HD~90~273 & ON7~V~((f))\\
CPD~-58~2611 & O6~V~((f))z & HD~12~993 & O6.5~V~((f))Nstr\\
\hline
    \end{tabular}
    \caption{Proposed new ON star. The table lists, for each case, the star proposed to be reclassified as ON and its current spectral classification, together with the same information for the reference star used for the direct comparison (ON when available, or the standard O6.5~III~(f) star HD~96~946 to compare with HD~190~864).}
    \label{tab_ON_rev} 
\end{table}

One particular case deserves additional discussion, as it is already classified as Nstr but we propose it should be reclassified as ON. For HD~12~993, we derive a N abundance comparable to that of HD~90~273, whereas a lower \Nab{} would be expected given its spectral type. Moreover, when compared to other O6.5~V stars in the sample, HD~12~993 shows a markedly higher nitrogen enrichment relative to its classification, further supporting its reclassification as ON. Consequently, CPD~-58~2611 should also be included in the ON group. According to our revised classification, ON stars form a clearly distinct group within our sample, characterized by \Nab{}\,$>\!8.45$~dex.

\end{appendix} 

\end{document}